\documentclass{aastex63}

\usepackage{natbib}
\usepackage{hyperref}
\received{December  xx, 2020}
\revised{XXXX}
\accepted{xxx}
\submitjournal{ApJ}
\shorttitle{Observational evidence from  CEMP stars}
\shortauthors{Purandardas and Goswami}
\graphicspath{{./}{figures/}}

\begin{document}

\title{ Observational evidence points at AGB stars as possible progenitors of CEMP-s \& r/s stars \footnote{ Based on data collected at Himalayan Chandra Telescope, IAO, Hanle, India and at Subaru Telescope, which is operated by the National Astronomical Observatory of Japan. The subaru data are retrieved from the JVO portal (\url{http://jvo.nao.ac.jp/portal/v2/}) operated by the NAOJ }}

\correspondingauthor{Aruna Goswami}
\email{aruna@iiap.res.in}
\author[0000-0001-5047-5950]{Meenakshi Purandardas}
\affiliation{Indian Institute of Astrophysics, Koramangala, Bangalore 560034, India;}
\affiliation{Department of physics, Bangalore university, Jnana Bharathi Campus, Karnataka 560056, India}

\author[0000-0002-8841-8641]{Aruna Goswami}
\affiliation{Indian Institute of Astrophysics, Koramangala, Bangalore 560034, India;}

\begin{abstract}
Origin of enhanced abundance of heavy elements observed in the surface chemical composition of carbon-enhanced metal-poor (CEMP) stars still remain poorly understood. Here, we present detailed abundance analysis  of seven CEMP stars based on high resolution (R${\sim}$ 50\,000) spectra  that reveal enough evidence of Asymptotic Giant Branch (AGB) stars being possible progenitors for these objects. For the objects HE0110$-$0406, HE1425$-$2052 and HE1428$-$1950,  we present for the first time a detailed abundance analysis.
Our sample is found to consists of one metal-poor ([Fe/H]$<$$-1.0$) and six very metal-poor ([Fe/H]$<$$-2.0$) stars with enhanced carbon and neutron-capture elements.   We have critically analysed the observed abundance ratios of [O/Fe], [Sr/Ba] and [hs/ls] and examined the possibility of AGB stars being possible progenitors.  The abundance of oxygen estimated in the programme stars are characteristics of  AGB progenitors except for HE1429$-$0551 and HE1447$+$0102. The estimated values of [Sr/Ba] and [hs/ls] ratios also support AGB stars as  possible progenitors. The locations of the programme stars in the absolute carbon abundance  A(C) vs. [Fe/H] diagram along with the Group I objects hint at binary nature of the object. We have studied  the chemical enrichment histories of the programme stars based on abundance ratios [Mg/C], [Sc/Mn] and [C/Cr]. Using [C/N] and $^{12}$C/$^{13}$C ratios we have examined if  any internal mixing had modified their surface chemical compositions. Kinematic analysis shows that the objects HE 0110$-$0406 and HE1447$+$0102 are  thick disk objects and the remaining five objects  belong to the halo population  of the Galaxy.
\end{abstract}

\section{Introduction}\label{sec:intro}

A substantial fraction of iron-deficient stars show enhancement of carbon \citep{rossi.1999ASPC..165..264R}. Studies of these group of stars are of special significance as their surface compositions are believed to hold the fossil records of the nucleosynthetic products of first stars. Such  metal-poor ([Fe/H] $<$ $-1.0$) stars with enhanced carbon ([C/Fe] $>$ 0.7, \cite{aoki.2007ApJ...655..492A}) are known as carbon-enhanced metal-poor (CEMP) stars.  Majority of the CEMP stars also exhibit enhancement of neutron-capture elements \citep{norris.1997ApJ...488..350N,barbuy.1997A&A...317L..63B,sneden.2003ApJ...591..936S}. CEMP stars are classified into different subgroups based on the abundance patterns of heavy elements \citep{beers&christlieb2005ARA&A..43..531B}.    

\par Initial classification of CEMP stars was based on the abundance patterns of heavy elements Ba and Eu \citep{beers&christlieb2005ARA&A..43..531B}. Subsequently, different authors \citep{jonsel.2006A&A...451..651J,masseron.2010A&A...509A..93M,abate.2016A&A...587A..50A,frebel.2018ARNPS..68..237F,hansen.2019A&A...623A.128H,partha.2021A&A...649A..49G} have used different criteria to classify the CEMP stars into various subclasses. A comprehensive and comparative study of the classification schemes can be found  in \cite{partha.2021A&A...649A..49G}. CEMP stars that are found to be enhanced in slow neutron-capture process (s-process) elements are called CEMP-s stars. The majority of CEMP stars belong to this group.  The observed enhancement of carbon and s-process elements in CEMP-s stars is thought to be originated due to mass transfer from a binary companion that has once gone through the AGB phase and synthesised the carbon and heavy elements. The radial velocity variations exhibited by these objects provide  strong evidence in favour of the binary mass transfer scenario \citep{McClure&Woodsworth1990ApJ...352..709M, preston&sneden2001AJ....122.1545P, hansenetal.2016A&A...588A...3H, jorrisen.2016A&A...586A.159J}. \cite{hansenetal.2016A&A...588A...3H} report that 82\% of CEMP-s stars in their sample exhibit radial velocity variations and are confirmed binaries. These authors  could not find any signatures of orbital motion for four of the CEMP-s stars in their sample and these stars appeared to be single or in a wide binary system. \cite{choplin.2017A&A...607L...3C} showed that the abundance patterns of three of these four CEMP-s stars match well with the yields of fast rotating massive stars.

\par CEMP stars that show over abundances of rapid neutron-capture process (r-process) elements are known as CEMP-r stars. CEMP-r stars are the rarest sub-group of CEMP stars and are found to be unlikely in a binary system \citep{hansen.2011ApJ...743L...1H}. \cite{hansen.2015ApJ...807..173H} suggest that the enhancement of r-process elements in CEMP-r stars is not attributed to any binary companion, but it might be inherited from the interstellar medium from which the star was formed.

\par CEMP stars that are enhanced in both s- and r- process elements are called CEMP-r/s stars. Many studies \citep{hill.2000A&A...353..557H,cohen.2003ApJ...588.1082C,qian.2003ApJ...588.1099Q,jonsel.2006A&A...451..651J}  have discussed various formation scenarios to explain the peculiar abundance patterns observed in CEMP-r/s stars. However none of these processes could re-produce the observed frequencies of CEMP-r/s stars except the intermediate neutron-capture process or i-process \citep{jonsel.2006A&A...451..651J,abate.2016A&A...587A..50A}. The i-process  occurs at a neutron density which is intermediate to the neutron densities required for s- and r-processes \citep{Dardelet:20150w,roederer.2016ApJ...821...37R,hampel.2016ApJ...831..171H,hampel.2019ApJ...887...11H}. \cite{hampel.2016ApJ...831..171H} could reproduce the observed frequencies of CEMP-r/s stars based on their model calculations of i-process nucleosynthesis. Many authors have showed that i-process can occur at different evolutionary stages of an AGB star \citep{cowan&rose1977ApJ...212..149C,campbell.2008A&A...490..769C,cristallo.2009PASA...26..139C,campbell.2010A&A...522L...6C,stancliffe.2011ApJ...742..121S,banerjee.2018MNRAS.480.4963B,clarkson.2018MNRAS.474L..37C} as well as in rapidly accreting white dwarf \citep{denissenkov.2017ApJ...834L..10D,denissencov.2019MNRAS.488.4258D} . According to this postulate, CEMP-r/s star is expected to be in a binary system and accreted the materials produced by its companion through i-process  which is supported by the discovery of many of the CEMP-r/s stars in binary systems \citep{barbuy.2005A&A...429.1031B,hansenetal.2016A&A...588A...3H}.  

\par Another group of CEMP stars that do not show any enhancement in heavy elements are called CEMP-no stars. As shown by many studies, the peculiar abundance patterns observed in  CEMP-no stars is attributed to the interstellar medium from which they are formed, is polluted by Spin stars \citep{meynet.2010A&A...521A..30M,chiappini.2013AN....334..595C} or Faint supernovae that underwent mixing and fall back \citep{umeda.2005ApJ...619..427U,nomoto.2013ARA&A..51..457N,tominaga.2014ApJ...785...98T} or metal-free massive stars \citep{heger.woosely.2010ApJ...724..341H}. Although most of the CEMP-no stars are found to be single objects \citep{norris.2013ApJ...762...28N}, a few of them are found to be confirmed binaries \citep{hansen.2016aA&A...586A.160H}. \cite{arentsen.2019A&A...621A.108A} suggest that the CEMP-no stars may be in a binary system with an extremely metal-poor companion that once passed through the AGB phase that had not produced any significant amount of s-process elements.

\par In this paper, we present the results from a high-resolution spectroscopic analysis of seven CEMP stars. The paper is organized as follows: section \ref{sec:pre-studies} presents a brief summary of the previous works on the programme stars. The selection of the objects and the source  of  spectra are described  in section \ref{sec:source-spectra}. In section \ref{sec:phtometric-temp}, we present the photometric temperature estimates of the programme stars. The methodology used for the estimation of atmospheric parameters is presented in section \ref{sec:spec-parameters}. The results of the abundance analysis are presented in section \ref{sec:abund-analysis}. Uncertainties in our abundance estimates are discussed in section \ref{sec:error_analysis}. The kinematic analysis is presented in section  \ref{sec:kinematic_analysis}. Interpretations of the results are discussed in section \ref{sec:discussions} and the conclusions are drawn in section \ref{sec:conclusions}. 
\section{Previous studies on the program stars: a brief summary  and novelty of this work}\label{sec:pre-studies}
Some aspects of the programme stars were addressed by different authors. Here we present a summary of their main results and the novelty of this work.\\
{\bf HE 0110$-$0406, and HE 1425$-$2052}\\
\cite{goswami.2005MNRAS.359..531G} and \cite{goswami.2010MNRAS.402.1111G} classified these objects as  potential CH star candidates based on low resolution (R$\sim$ 1300) spectroscopic analysis. \cite{beers.2007ApJS..168..128B} presented the results of broadband BVR$_{C}$I$_{C}$CCD photometry for HE 1425$-$2052.
We present for the first time an abundance analysis for the object HE 1425$-$2052 using high-resolution spectra. For HE 0110$-$0406, we have performed for the first time a high-resolution spectroscopic analysis and presented some results in our earlier two papers, \cite{purandardas.2019bBSRSL..88..207P} and \cite{purandardas.2020JApA...41...36P}. In this work, we have examined the possible progenitor of this object based on the estimated elemental abundance ratios.\\
{\bf HE 1428$-$1950}\\
\cite{beers.2007ApJS..168..128B} presented  photometric analysis of this object. From low-resolution spectroscopic analysis  \cite{goswami.2010MNRAS.402.1111G} showed that  this object exhibits the spectral properties of a C-R star. The spectrum of this object was found to match well with the spectrum of the well known C-R star RV Sct, with the CH band around 4300 {\rm \AA}, slightly stronger than that in the RV Sct star.  The object also showed stronger Ca I line at 4226 {\rm \AA}, and a stronger feature due to Na D1 and D2 than usually seen in the spectra of CH stars.  \cite{placco.2011AJ....142..188P} have reported the atmospheric parameters for this object based on the analysis of low-resolution (R $\sim$ 2000) spectra. \cite{kennedy.2011AJ....141..102K} have presented the atmospheric parameters, metallicity, carbon and oxygen abundances of this star using the low- resolution optical (R$\sim$ 2000) and  medium-resolution (R $\sim$ 3000) near IR spectra. Polarimetric studies of \cite{goswami.2013A&A...549A..68G} reported  significant polarization for this object  over different colours of BVRI with percentage of polarization in the range 0.2 to 0.9 that  indicates the presence of a circumstellar envelope for this object. \cite{munari.2014AJ....148...81M} reported the effective temperature and reddening of this star using the AAVSO Photometric All-Sky Survey (APASS) data. Although some properties of this object have been addressed in many studies, results based on high-resolution spectroscopic analysis are missing. We present for the first time a detailed abundance analysis for this object using high-resolution spectra.\\
{\bf HE 1429$-$0551 and HE 1447+0102 }\\
Low-resolution spectroscopic analysis of HE 1429$-$0551 and HE 1447+0102 by \cite{goswami.2005MNRAS.359..531G} and \cite{goswami.2010MNRAS.402.1111G} showed that these objects are potential CH star candidate. \cite{beers.2007ApJS..168..128B} reported the B, V and R magnitudes for these objects. For both the stars, \cite{aoki.2007ApJ...655..492A} presented the abundance analysis results for C, N, Na, Mg, Ca, Ti, and Ba using the high-resolution (R$\sim$ 50000) spectra obtained using the High Dispersion Spectrograph (HDS) of the Subaru telescope. Polarimetric studies of \cite{goswami.2010ApJ...722L..90G} in the V-band, reported percentage of polarization ${\sim}$  1.3 $\pm$ 0.06 for HE 1429$-$0551 that indicates the existence of a circumstellar envelop for this object.  \cite{bisterzo.2011MNRAS.418..284B} grouped these objects as a CEMP-s II stars that fall in the category of CEMP-s star with [hs/ls] $\geq$ 1.5. They used the value for [hs/ls] from \cite{aoki.2007ApJ...655..492A} for their classification. Based on their model calculations, they claimed that the observed abundance patterns in HE 1429$-$0551 and HE 1447+0102 may be due to mass transfer from an AGB companion with a mass in the range 1.4 - 2 M$_{\odot}$ and 1.5 - 2 M$_{\odot}$ respectively. \cite{jorrisen.2016A&A...586A.159J} studied the radial velocity variations exhibited by HE 1429$-$0551 and showed that the object does not exhibit any signs of orbital motion and it requires more observations to confirm its binarity. \cite{karinkuzhi.2021A&A...645A..61K} report the abundances of a few light and heavy elements for HE 1429$-$0551 based on an analysis of high-resolution (R $\sim$ 86 000) spectrum obtained with HERMES spectrograph mounted on the Mercator telescope. In this work, we have reported elemental  abundances of two other elements Sc and V for HE 1429$-$0551 and elemental abundances of O, Sc, V, Cr, Mn, Ni, Sr, La, Ce, Pr, Nd, Sm, and Eu for HE 1447+0102 that are not available in literature. We have attempted to find the possible progenitors of these objects based on the observed abundance patterns. We have also checked the possibility of any internal mixing  based on [C/N] and $^{12}$C/$^{13}$C ratios which was not addressed in any of the previous works.\\
{\bf HE 1523$-$1155, and HE 1528$-$0409}\\
\cite{goswami.2005MNRAS.359..531G} classified these objects as  potential CH star candidates based on the low-resolution spectroscopic analysis. \cite{beers.2007ApJS..168..128B} reported B, V and R magnitudes for these object. \cite{aoki.2007ApJ...655..492A} reported the abundances of C, N, Na, Mg, Ca, Ti, and Ba in these objects based on high-resolution spectroscopic analysis. Polarimetric studies by \cite{goswami.2010ApJ...722L..90G} showed a percentage of polarization ${\sim}$ 0.85 $\pm$ 0.08 in V-band for these objects that indicates presence of a circumstellar envelope.  \cite{bisterzo.2011MNRAS.418..284B} classified these objects as CEMP-s II stars with the observed abundance patterns attributed to an AGB companion with a mass in the range 1.5 - 2.0 M$_{\odot}$. \cite{hansenetal.2016A&A...588A...3H} showed that the object HE 1523$-$1155 is a binary star based on the radial velocity variations exhibited by this object. They claimed that more radial velocity monitoring of the object is required to determine the final orbital parameters with confidence. In this work, we have presented the estimates and analyses of a few more elements such as O, Sc, V, Cr, Ni, Zn, Sr, Y, La, Ce, Pr, Nd, Sm, and Eu in these objects, that are not available in literature.  \\

\section{Object selection and source of spectra}\label{sec:source-spectra}
\begin{figure}
\centering
\includegraphics[width=11cm,height=9cm]{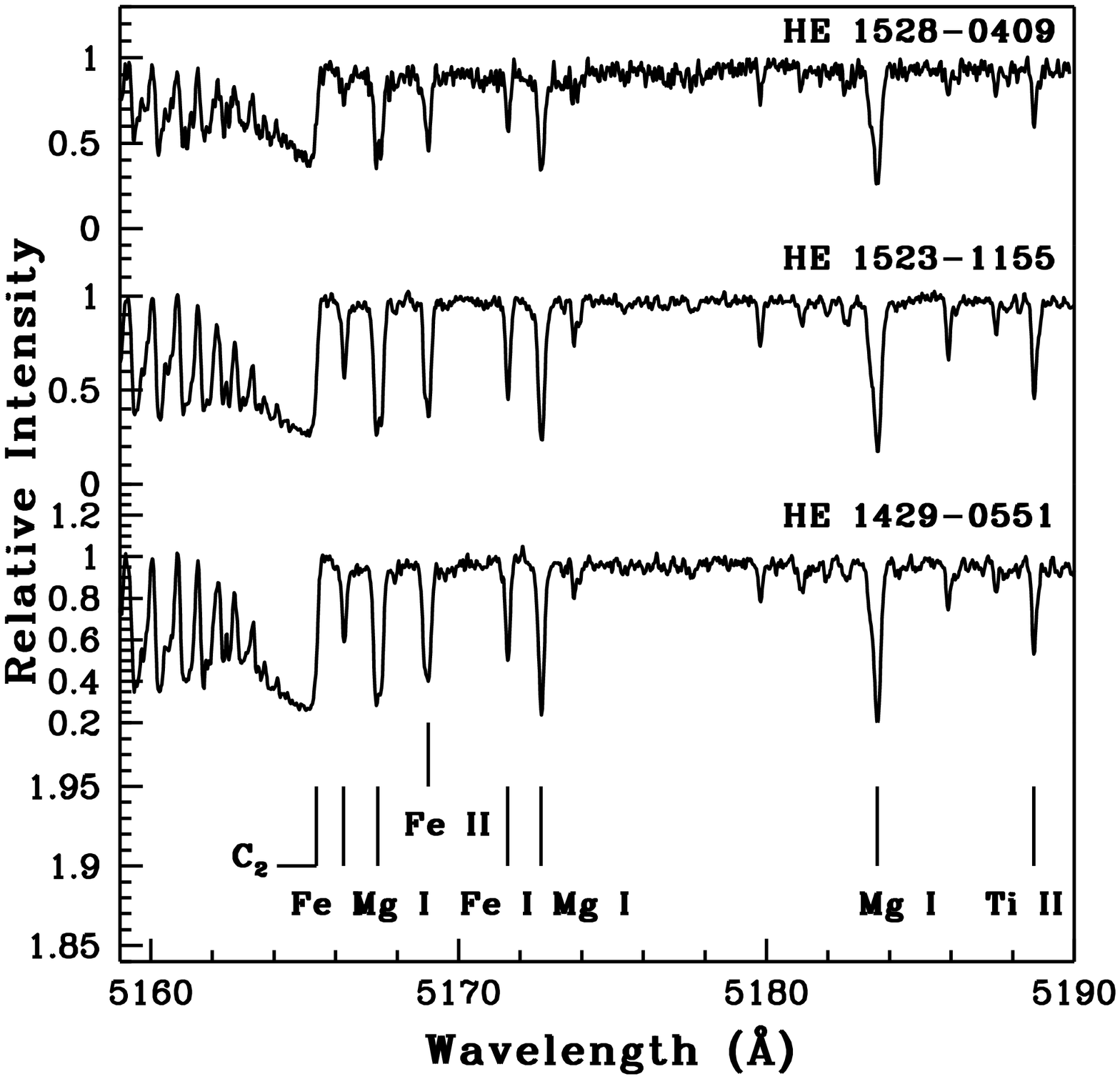}
\includegraphics[width=11cm,height=9cm]{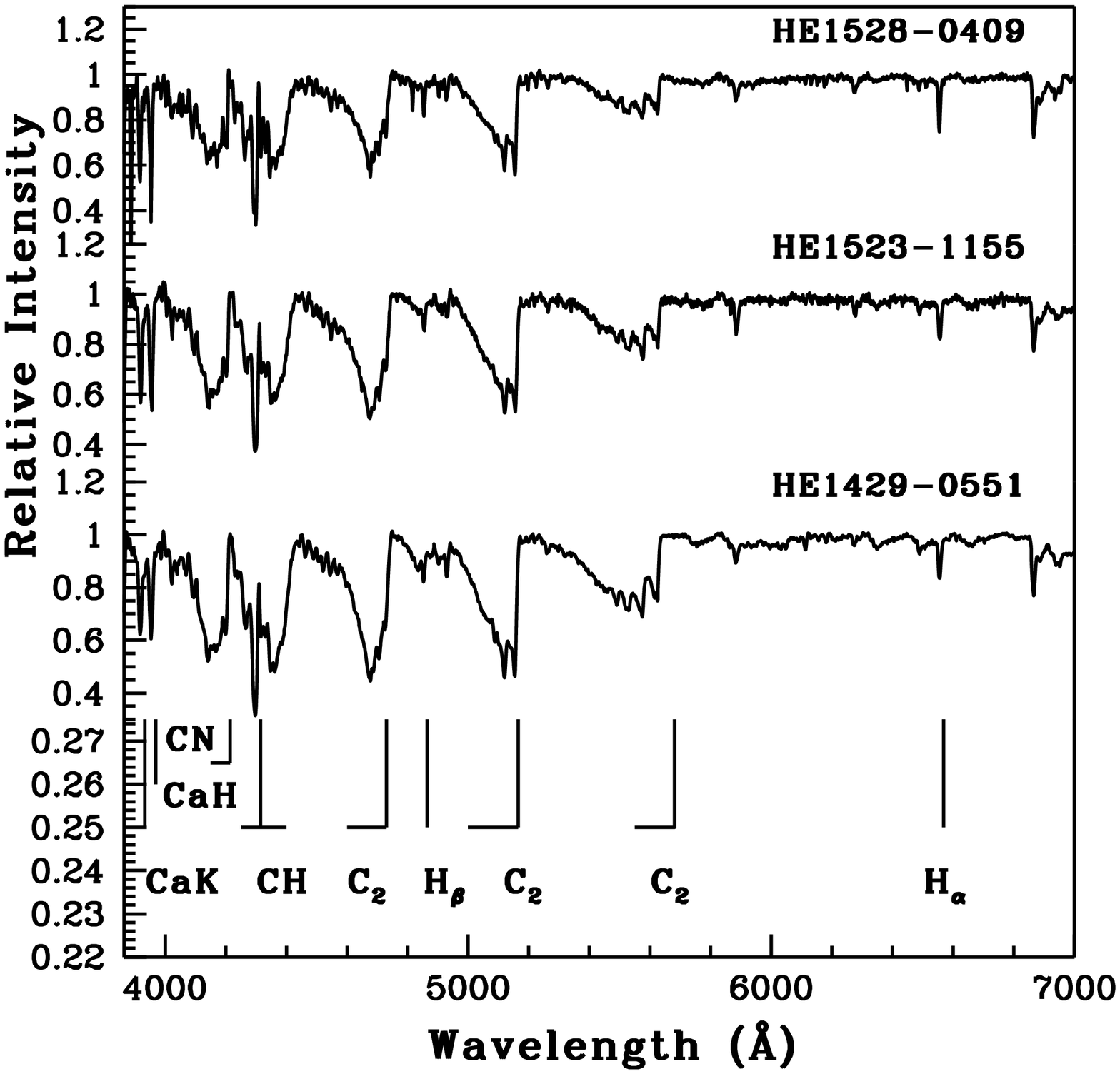}
\caption{ Upper panel: Examples of high resolution spectra of the programme stars  in the wavelength region 5160 to 5190 {\rm \AA}.
Lower panel:  Examples of low resolution spectra of the programme stars in the wavelength region 3860 to 7000 {\rm \AA}}. 
\label{fig1}
\end{figure}

Objects are selected from the lists of potential CH star candidates of 
\cite{goswami.2005MNRAS.359..531G} and \cite{goswami.2010MNRAS.402.1111G} for a follow-up detailed high resolution spectroscopic analysis to confirm their classification as well as to understand  the characteristic properties of their surface chemical composition. In  Figure \ref{fig1}  (Lower panel) a few sample spectra at low-resolution obtained using the Himalayan Faint Object Spectrograph Camera (HFOSC) and grism 7  covering  the wavelength regions from 3860 - 7000 {\rm \AA} are shown. The spectral  resolution ($\lambda/\delta\lambda$) is   $\sim$ 1330.  

\par For the object HE 0110$-$0406, the high resolution ($\lambda/\delta\lambda \sim 60,000 $) spectrum was obtained using the high resolution fiber fed  Hanle Echelle SPectrograph (HESP) attached to the Himalayan Chandra Telescope (HCT). The wavelength coverage of the spectra spans from 3530-9970 {\rm \AA}. The spectrograph operates at a resolution of 60 000 with a slicer, and a resolution of 30 000 without slicer. A CCD with 4096$\times$4096 pixels of 15 micron size is used to record the spectrum. The~high-resolution~spectra of the remaining programme stars were retrieved from the JVO portal \url{http://jvo.nao.ac.jp/portal/v2/} operated by the National Astronomical Observatory of Japan (NAOJ) where the reduced and wavelength calibrated spectra are publicly available. These spectra were obtained using the High Dispersion Spectrograph (HDS) of the 8.2 m Subaru Telescope with a resolution of $\sim$ 50 000 and a wavelength coverage from 4100 to 6850{\rm \AA}, with a small gap between 5440 and 5520{\rm \AA} due to the physical separation between the two EEV CCDs with 2048$\times$4096 pixels with two by two on-chip binning. The signal to noise ratio of the spectra of the programme stars range from $\sim$ 53 - 67 per resolution element measured at 6000 \AA. The basic data of the programme stars are given in Table \ref{table1}. Figure \ref{fig1} (Upper panel) shows a few examples of sample spectra at high resolution.  

{\footnotesize
\begin{table*}
\caption{\bf Basic data for the programme stars}
\resizebox{1\textwidth}{!}{\begin{tabular}{lccccccccccc}
\hline
Star      &RA$(2000)$       &Dec.$(2000)$    &B       &V       &J        &H        &K  &Exposure   & S/N      & Date of obs. & Source \\
          &                 &                &        &        &         &         &   &  (seconds) & (at 6000\AA)    &        & of spectra\\
\hline
HE 0110-0406& 01 12 37.16   & $-03$ 50 29.15 & 13.77  & 12.48  & 10.52   & 9.98    & 9.86  & 2700 & 66 &24-12-2017  & HCT/HESP\\ 
HE1425-2052 & 14 28 39.54   & $-21$ 06 04.78 & 13.97  & 12.70  & 10.04   & 9.45    & 9.27  & 600  & 64 &25-05-2003  & Subaru/HDS\\ 
HE1428-1950 & 14 30 59.39   & $-20$ 03 41.91 & 13.28  & 12.04  & 9.98    & 9.47    & 9.32  & 600  & 58 &25-05-2003  & Subaru/HDS\\
HE1429-0551 & 14 32 31.29   & $-06$ 05 00.20 & 14.01  & 12.61  & 10.73   & 10.27   & 10.07 & 600  & 65 &25-05-2003  & Subaru/HDS\\ 
HE1447+0102 & 14 50 15.04   & 00    50 15.09 & 15.63  & 14.61  & 13.21   & 12.76   & 12.68 & 600  & 57 &26-05-2003  & Subaru/HDS\\
HE1523-1155 & 15 26 41.04   & $-12$ 05 42.66 & 14.57  & 13.22  & 11.37   & 10.85   & 10.75 & 600  & 67 &26-05-2003  & Subaru/HDS\\
HE1528-0409 & 15 30 54.30   & $-04$ 19 40.36 & 16.00  & 14.76  & 12.94   & 12.45   & 12.36 & 600  & 53 &26-05-2003  & Subaru/HDS\\
\hline
\end{tabular}}
\label{table1}
\end{table*}
}

\section{Photometric temperatures}\label{sec:phtometric-temp}
We have determined the photometric temperatures for our programme stars using the calibration equations of \cite{alonso.1999A&AS..140..261A} following the procedures as described in \citet{goswami.2006MNRAS.372..343G,goswami.2016MNRAS.455..402G}.  Interstellar extinctions  for objects with b $<$ 50 are determined using  the formulae from  Chen et al. (1998). The estimated values of reddening for our programme stars range from 0 - 0.04 and are negligible. J, H, and K magnitudes of the objects are taken from SIMBAD data base which came from 2MASS \citep{cutri.2003yCat.2246....0C}. Photometric temperatures are used as the initial guess for the determination of spectroscopic temperatures of the programme stars. The photometric temperatures obtained for our programme stars are presented in Table \ref{table2}.

{\footnotesize
\begin{table*}
\caption{\bf Temperatures from  photometry }
\resizebox{\textwidth}{!}{\begin{tabular}{llllllllllllll}
\hline                      
Star name   & $T_eff$ &$T_eff $ &$T_eff $&  $T_eff$ & $T_eff$ &$T_eff $ &  $T_eff$ & $T_eff$  & $T_eff $ & $T_eff$  & $T_eff$ & $T_eff$ & Spectroscopic\\
            &         &($-1.5$)  &($-2.0)$& ($-2.5$)& [Fe/H] &($-1.5$) &  ($-2.0$)& ($-2.5$)& [Fe/H] &($-1.5$)& ($-2.0$) & [Fe/H] &estimates  \\
            & (J$-$K) & (J$-$H) &(J$-$H) & (J$-$H) & (J-H) &(V$-$K)  &  (V$-$K) & (V$-$K) & (V$-$K) & (B$-$V)& (B$-$V)& (B$-$V) &  \\
\hline
HE 0110$-$0406& 4657.90 & 4756.88 & 4733.47 & 4688.11 & 4759.92 & 4447.21 & 4446.24&4451.10 & 4449.23 & 4162.19 & 4167.47 & 4166.76 &4670 \\
HE 1425$-$2052& 4358.16 & 4526.41 &4503.99& 4486.04 & 4497.10 & - &- & -  &  - & 4189.81 & 4193.06  & 4196.62 & 4300\\
HE 1428$-$1950& 4620.68 & 4820.74&4797.06& 4750.81 & 4788.78 &4365.58&4364.03  & 4368.09 & 4364.44 & 4231.34& 4231.51  &  4234.86 &4500\\
HE 1429$-$0551& 4626.35 & 5064.78&5040.04& 4990.39 & 4994.05 & 4507.99&4507.46  & 4512.92 & 4512.42 &- & -  &  - &4940 \\ 
HE 1447+0102& 5084.56   &5133.68 &5108.64& 5057.99 & 5082.30 &5136.28&5144.59  & 5160.14 & 5152.73 &4519.41& 4496.74 &  4496.38 & 5220 \\ 
HE 1523$-$1155& 4755.96 & 4792.56&4769.00& 4723.15 & 4730.87 &4561.47&4561.32  & 4567.31 &  4566.09  &4099.98& 4109.71  &  4136.83 & 4780\\
HE 1528$-$0409& 4872.38 & 4941.53&4917.33&  4869.41 & 4884.08 &4624.51 &4621.44 &  4624.16 &  4622.89 & 4222.72 & 4223.54  &  4236.76 & 5250 \\
\hline      
\end{tabular}}

\textit{Note.} The numbers in the parenthesis below $T_{eff}$ indicate the metallicity values at which the temperatures are 
calculated. The temperatures calculated using the adopted metallicity of the stars are presented in columns 6, 10 and 13. Temperatures are given in Kelvin. 
\label{table2}
\end{table*}
}

\section{Spectroscopic stellar parameters}\label{sec:spec-parameters}
Radial velocities of the programme stars are measured using the shift in the wavelength for a large number of clean and unblended lines in their spectra. The estimated radial velocities range from $-151.5$ to 121.2 Km s$^{-1}$ (Table \ref{table3}). Except for HE 1447+0102 and HE 1528$-$0409 for which the radial velocity information are not available in literature, our estimates of radial velocities of the remaining  programme stars show a difference of $\sim$ 1.6 – 11.8 km s$^{-1}$ from the corresponding literature values which may be a clear indication of the stars being in binary systems.

{\footnotesize
\begin{table*}
\caption{\bf Radial velocities of the programme stars. }
\begin{tabular}{lcc}
\hline
Star            & V$_{r}$             & Reference \\
                & (Kms$^{-1}$)          & \\
\hline
HE 0110-0406  	&	 $-44.4$$\pm$3.80 	& 1	\\ 
                &    $-32.8$$\pm$2.30   & 2 \\
HE 1425$-$2052 	&	 121.2$\pm$0.22  	& 1	\\
                &    133$\pm$1.59       & 2 \\
HE 1428$-$1950 	&	 58.8$\pm$0.44      & 1 \\ 	  
                &	 62.1$\pm$0.82      & 2 \\
                &   61.9$\pm$3.58 &  3 \\
HE 1429$-$0551 	&	 $-48.6$$\pm$0.11   & 1 \\ 	
                &	 $-43.1$$\pm$0.88   & 2 \\
                &   $-44.9\pm$0.11 & 4 \\
HE 1447+0102  	&	 $-61.8$$\pm$0.10 	& 1 \\
                &    $-61.2$$\pm$0.21   &  4\\
HE 1523$-$1155 	&	 $-40.5$$\pm$0.08   & 1 \\ 	
                &	 $-38.9$$\pm$3.50   & 2\\
                &   $-45.0$$\pm$0.12  & 4\\
HE 1528$-$0409 	&	 $-151.5$$\pm$0.13 & 1\\
                &    $-157.5$$\pm$0.25 & 4 \\

\hline
\end{tabular}

1. Our work, 2. \cite{gaia.2018A&A...616A...1G}, 3. \cite{kunder.2017AJ....153...75K}, 4.  \cite{aoki.2007ApJ...655..492A}
\label{table3}
\end{table*}

}

\par
\begin{figure}
\centering
\includegraphics[width=11cm,height=10cm]{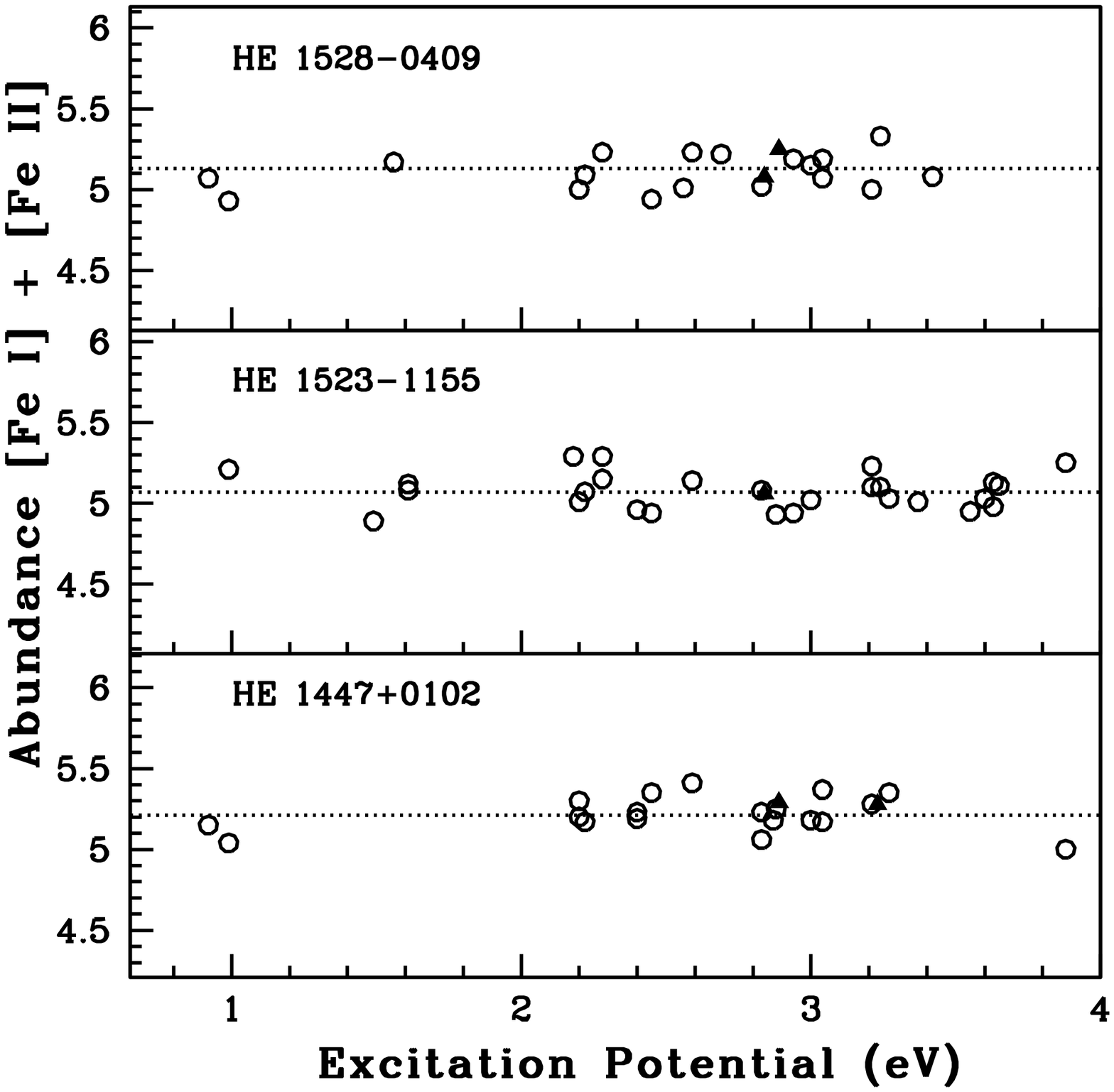}
\includegraphics[width=11cm,height=10cm]{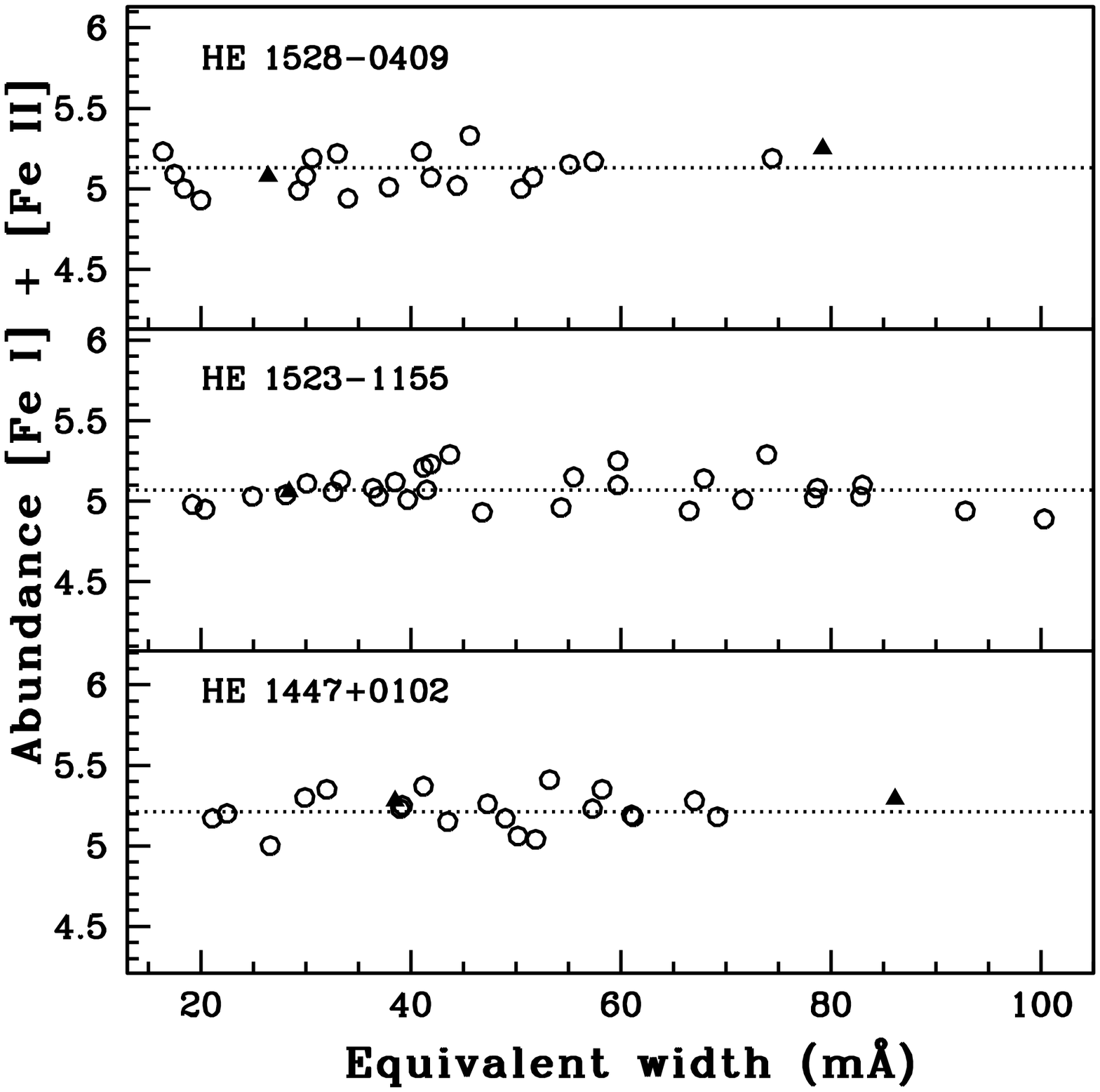}
\caption{ The iron abundances of programme stars as a function of excitation potential (upper panel) and equivalent widths (lower panel). The open circles indicate Fe I lines and solid triangles represent Fe II lines.}
\label{fig2}
\end{figure}

 Atmospheric parameters the effective temperature T$_{eff}$, surface gravity, log\,g and microturbulent velocity ${\zeta}$  of the programme stars are determined from the measured equivalent widths of a set of clean lines due to Fe I and Fe II (Table \ref{table4}). Effective temperature is taken as the value obtained when the trend between the abundances of Fe I and Fe II and the corresponding excitation potentials gives a zero slope. At this temperature, microturbulent velocity is fixed to be that value for which there is no trend between the abundances derived from the Fe I and Fe II lines and the reduced equivalent widths. Under this temperature and microturbulent velocity, the surface gravity, log g is taken to be that value corresponding to which the abundances of Fe I and Fe II are nearly the same. The abundances of Fe I and Fe II for the programme stars as a function of excitation potential and as a function of equivalent widths are shown in Figure \ref{fig2}. 

{\footnotesize
\begin{table*}
\caption{\bf Equivalent widths (in m\r{A}) of Fe lines used for deriving atmospheric parameters.}
\resizebox{\textwidth}{!}{\begin{tabular}{ccccccccccc}
\hline         
Wavelength(\r{A}) & Element & $E_{low}$(eV) & log gf & HE 0110-0406 & HE 1425-2052 & HE 1428-1950 &	HE 1429-0551 & HE 1447+0102 & HE 1523-1155		& HE 1528-0409	\\
\hline																				-	
4139.927	&	Fe I	&	0.990	&	-3.629	&	-	&	-	&	-	&	-	&	-	&	41.2(5.21)	&	20.0(5.42)	\\
4337.046	&		&	1.557	&	-1.695	&	-	&	-	&	-	&	78.1(4.86)	&	-	&	-	&	-	\\
4422.567	&		&	2.845	&	-1.110	&	-	&	96.8(5.10)	&	-	&	-	&	-	&	-	&	-	\\
4445.471	&		&	0.087	&	-5.441	&	-	&	60.0(5.30)	&	65.3(5.47)	&	-	&	-	&	-	&	-	\\
4446.833	&		&	3.686	&	-1.330	&	-	&	-	&	-	&	-	&	-	&	78.7(5.08)	&	-	\\
4466.551	&		&	2.832	&	-0.590	&	-	&	-	&	135.7(5.15)	&	70.7(5.07)	&	50.2(4.97)	&	-	&	-	\\
4484.219	&		&	3.603	&	-0.720	&	-	&	87.9(5.56)	&	55.5(5.19)	&	-	&	-	&	24.9(5.03)	&	-	\\
4531.147	&		&	1.485	&	-2.155	&	-	&	-	&	-	&	62.2(4.87)	&	-	&	-	&	-	\\
4566.514	&		&	3.301	&	-2.250	&	24.6(6.07)	&	-	&	-	&	-	&	-	&	-	&	-	\\
4595.358	&		&	3.302	&	-1.720	&	60.9(6.16)	&	-	&	34.2(5.50)	&	-	&	-	&	-	&	-	\\
4630.121	&		&	2.279	&	-2.600	&	-	&	-	&	60.6(5.44)	&	-	&	-	&	-	&	-	\\
4632.911	&		&	1.608	&	-2.913	&	-	&	-	&	77.7(5.08)	&	-	&	-	&	38.5(5.12)	&	-	\\
4643.463	&		&	3.654	&	-1.290	&	-	&	-	&	27.0(5.37)	&	-	&	-	&	-	&	-	\\
4690.136	&		&	3.686	&	-1.680	&	31.7(6.09)	&	-	&	-	&	-	&	-	&	-	&	-	\\
4745.800	&		&	3.654	&	-0.790	&	-	&	-	&	77.3(5.55)	&	-	&	-	&	-	&	-	\\
4789.655	&		&	3.547	&	-0.840	&	-	&	-	&	-	&	25.4(5.21)	&	-	&	20.4(4.95)	&	-	\\
4871.317	&		&	2.868	&	-0.410	&	-	&	166.7(5.36)	&	160.8(5.27)	&	89.0(5.23)	&	69.2(5.18)	&	-	&	-	\\
4882.144	&		&	3.417	&	-1.640	&	62.6(6.21)	&	-	&	-	&	-	&	-	&	-	&	-	\\
4903.308	&		&	2.882	&	-1.080	&	-	&	104.4(5.14)	&	131.1(5.51)	&	43.2(5.04)	&	39.2(5.25)	&	46.8(4.93)	&	-	\\
4924.770	&		&	2.279	&	-2.220	&	-	&	-	&	120.2(5.69)	&	-	&	-	&	43.7(5.29)	&	26.4(5.23)	\\
4939.686	&		&	0.859	&	-3.340	&	-	&	-	&	164.4(5.50)	&	-	&	-	&	-	&	-	\\
4950.104	&		&	3.417	&	-1.670	&	58.1(6.17)	&	-	&	-	&	-	&	-	&	-	&	-	\\
4967.890	&		&	4.191	&	-0.622	&	65.0(6.15)	&	-	&	-	&	-	&	-	&	-	&	-	\\
4985.251	&		&	3.928	&	-0.485	&	-	&	83.6(5.65)	&	-	&	-	&	-	&	-	&	-	\\
4994.129	&		&	0.915	&	-3.080	&	-	&	-	&	-	&	-	&	43.5(5.15)	&	-	&	-	\\
5001.862	&		&	3.881	&	0.010	&	-	&	104.6(5.35)	&	-	&	37.2(4.96)	&	26.6(4.96)	&	59.7(5.25)	&	-	\\
5006.117	&		&	2.832	&	-0.767	&	-	&	-	&	-	&	-	&	57.3(5.23)	&	-	&	44.4(5.02)	\\
5028.126	&		&	3.573	&	-1.474	&	-	&	-	&	23.8(5.35)	&	-	&	-	&	-	&	-	\\
5051.634	&		&	0.915	&	-2.795	&	-	&	-	&	-	&	75.2(4.95)	&	-	&	-	&	51.6(5.07)	\\
5079.224	&		&	2.198	&	-2.067	&	-	&	-	&	-	&	38.8(5.13)	&	29.9(5.30)	&	-	&	-	\\
5109.650	&		&	4.301	&	-0.980	&	33.7(6.12)	&	-	&	-	&	-	&	-	&	-	&	-	\\
5166.281	&		&	0.000	&	-4.195	&	-	&	-	&	-	&	68.3(5.07)	&	47.3(5.26)	&	82.8(5.03)	&	29.3(4.99)	\\
5171.595	&		&	1.485	&	-1.793	&	-	&	-	&	-	&	96.7(5.03)	&	-	&	100.3(4.89)	&	-	\\
5192.343	&		&	2.998	&	-0.521	&	-	&	-	&	156.1(5.37)	&	64.9(4.96)	&	-	&	-	&	55.1(5.15)	\\
5194.941	&		&	1.557	&	-2.090	&	-	&	-	&	176.1(5.29)	&	88.1(5.24)	&	-	&	-	&	57.4(5.17)	\\
5195.468	&		&	4.221	&	-0.002	&	-	&	-	&	83.3(5.49)	&	-	&	-	&	28.1(5.04)	&	-	\\
5198.711	&		&	2.222	&	-2.135	&	-	&	130.5(5.60)	&	-	&	39.1(5.23)	&	21.1(5.17)	&	41.5(5.07)	&	27.5(5.09)	\\
5215.179	&		&	3.266	&	-0.933	&	106.5(5.99)	&	76.6(5.14)	&	-	&	32.9(5.11)	&	32.0(5.35)	&	36.9(5.03)	&	-	\\
5217.389	&		&	3.211	&	-1.097	&	113.8(6.22)	&	109.7(5.61)	&	-	&	-	&	-	&	41.9(5.23)	&	-	\\
5226.862	&		&	3.038	&	-0.667	&	-	&	-	&	-	&	42.9(4.76)	&	49.0(5.17)	&	-	&	41.9(5.07)	\\
5232.939	&		&	2.940	&	-0.190	&	-	&	161.7(5.05)	&	178.4(5.28)	&	86.6(4.97)	&	-	&	92.8(4.94)	&	74.4(5.19)	\\
5242.491	&		&	3.634	&	-0.840	&	-	&	-	&	82.4(5.58)	&	-	&	-	&	19.2(4.98)	&	-	\\
5247.049	&		&	0.087	&	-4.946	&	-	&	153.5(5.70)	&	112.2(5.30)	&	26.6(5.23)	&	-	&	32.6(5.06)	&	-	\\
5250.645	&		&	2.198	&	-2.050	&	-	&	-	&	-	&	27.0(4.87)	&	-	&	-	&	18.4(5.00)	\\
5253.461	&		&	3.283	&	-1.670	&	-	&	19.9(5.06)	&	-	&	-	&	-	&	-	&	-	\\
5263.305	&		&	3.266	&	-0.970	&	121.2(6.29)	&	-	&	-	&	-	&	-	&	-	&	-	\\
5266.555	&		&	2.998	&	-0.490	&	-	&	156.5(5.35)	&	160.9(5.39)	&	74.0(5.08)	&	61.2(5.18)	&	78.4(5.02)	&	-	\\
5281.790	&		&	3.038	&	-1.020	&	-	&	122.7(5.46)	&	-	&	50.8(5.25)	&	41.2(5.37)	&	-	&	30.6(5.19)	\\
5283.621	&		&	3.241	&	-0.630	&	-	&	-	&	-	&	-	&	-	&	59.7(5.10)	&	45.6(5.33)	\\
5288.528	&		&	3.695	&	-1.670	&	49.2(6.33)	&	-	&	-	&	-	&	-	&	-	&	-	\\
5307.360	&		&	1.608	&	-2.987	&	-	&	95.0(5.18)	&	-	&	27.2(5.12)	&	-	&	36.4(5.08)	&	-	\\
5324.178	&		&	3.211	&	-0.240	&	-	&	-	&	175.6(5.63)	&	63.4(4.88)	&	67.0(5.28)	&	83.0(5.10)	&	50.5(5.00)	\\
5339.928	&		&	3.626	&	-0.680	&	-	&	-	&	-	&	30.6(5.21)	&	-	&	33.3(5.13)	&	-	\\
5373.698	&		&	4.473	&	-0.860	&	45.4(6.39)	&	-	&	-	&	-	&	-	&	-	&	-	\\
5383.369	&		&	4.313	&	0.500	&	123.6(6.15)	&	-	&	-	&	-	&	-	&	-	&	-	\\
5389.479	&		&	4.415	&	-0.410	&	67.9(6.21)	&	-	&	-	&	-	&	-	&	-	&	-	\\
5424.069	&		&	4.320	&	0.520	&	123.0(6.12)	&	-	&	-	&	-	&	-	&	-	&	-	\\
5506.778	&		&	0.990	&	-2.797	&	-	&	-	&	-	&	-	&	51.9(5.04)	&	-	&	-	\\
5554.882	&		&	4.548	&	-0.440	&	-	&	24.6(5.52)	&	-	&	-	&	-	&	-	&	-	\\
5569.618	&		&	3.417	&	-0.540	&	-	&	-	&	-	&	-	&	-	&	-	&	30.0(5.08)	\\
5576.090	&		&	3.430	&	-1.000	&	-	&	97.9(5.63)	&	-	&	-	&	-	&	-	&	-	\\
5586.756	&		&	3.368	&	-0.210	&	-	&	142.1(5.32)	&	130.4(5.12)	&	64.8(5.04)	&	-	&	71.6(5.01)	&	-	\\
5701.545	&		&	2.559	&	-2.216	&	100.6(6.19)	&	-	&	-	&	-	&	-	&	-	&	-	\\
6024.049	&		&	4.548	&	-0.120	&	78.7(6.20)	&	54.9(5.66)	&	-	&	-	&	-	&	-	&	-	\\
6136.615	&		&	2.453	&	-1.400	&	-	&	161.0(5.40)	&	166.2(5.43)	&	69.2(5.17)	&	58.2(5.35)	&	66.5(4.94)	&	34.0(4.94)	\\
6136.993	&		&	2.198	&	-2.950	&	81.2(6.12)	&	-	&	-	&	-	&	-	&	-	&	-	\\
6137.694	&		&	2.588	&	-1.403	&	145.9(6.06)	&	161.9(5.60)	&	147.5(5.40)	&	-	&	53.2(5.41)	&	67.9(5.14)	&	41.0(5.23)	\\
6180.203	&		&	2.727	&	-2.780	&	57.4(6.30)	&	-	&	-	&	-	&	-	&	-	&	-	\\
\hline
\end{tabular}}
\label{table4}
\end{table*}
}

{\footnotesize
\begin{table*}
\resizebox{\textwidth}{!}{\begin{tabular}{ccccccccccc}
\hline         
Wavelength(\r{A}) & Element & $E_{low}$(eV) & log gf & HE 0110-0406 & HE 1425-2052 & HE 1428-1950 &	HE 1429-0551 & HE 1447+0102 & HE 1523-1155		& HE 1528-0409	\\
\hline	
6219.279	&		&	2.198	&	-2.433	&	126.1(6.24)	&	-	&	-	&	-	&	-	&	-	&	-	\\
6230.726	&		&	2.559	&	-1.281	&	167.3(6.24)	&	150.6(5.29)	&	-	&	52.8(4.90)	&	-	&	-	&	37.9(5.01)	\\
6246.317	&		&	3.603	&	-0.960	&	108.1(6.29)	&	-	&	-	&	-	&	-	&	-	&	-	\\
6252.554	&		&	2.404	&	-1.687	&	-	&	154.8(5.53)	&	163.0(5.59)	&	40.3(4.92)	&	39.0(5.23)	&	54.3(4.96)	&	-	\\
6301.498	&		&	3.654	&	-0.745	&	-	&	60.4(5.20)	&	67.2(5.24)	&	-	&	-	&	30.1(5.11)	&	-	\\
6335.328	&		&	2.198	&	-2.230	&	-	&	109.9(5.28)	&	123.3(5.41)	&	21.1(4.83)	&	22.5(5.20)	&	39.7(5.01)	&	-	\\
6411.647	&		&	3.654	&	-0.820	&	99.1(6.07)	&	-	&	-	&	-	&	-	&	-	&	-	\\
6421.349	&		&	2.278	&	-2.027	&	146.6(6.22)	&	-	&	-	&	39.8(5.09)	&	-	&	55.5(5.15)	&	-	\\
6430.844	&		&	2.176	&	-2.006	&	155.1(6.19)	&	-	&	-	&	-	&	-	&	73.9(5.29)	&	-	\\
6494.980	&		&	2.404	&	-1.273	&	-	&	-	&	-	&	59.4(4.79)	&	61.0(5.19)	&	-	&	-	\\
6575.019	&		&	2.588	&	-2.820	&	-	&	39.9(5.61)	&	-	&	-	&	-	&	-	&	-	\\
6593.871	&		&	2.433	&	-2.422	&	112.3(6.27)	&	93.3(5.60)	&	-	&	-	&	-	&	-	&	-	\\
6677.989	&		&	2.692	&	-1.470	&	-	&	-	&	-	&	-	&	-	&	-	&	33.0(5.22)	\\
4491.405	&	Fe II	&	2.856	&	-2.700	&	-	&	-	&	65.0(5.53)	&	-	&	-	&	-	&	-	\\
4508.288	&		&	2.856	&	-2.210	&	-	&	45.2(5.30)	&	76.5(5.21)	&	41.0(4.97)	&	-	&	-	&	-	\\
4515.339	&		&	2.844	&	-2.480	&	-	&	43.5(5.53)	&	-	&	-	&	-	&	28.4(5.06)	&	26.4(5.08)	\\
4731.453	&		&	2.891	&	-3.360	&	47.1(6.12)	&	-	&	-	&	-	&	-	&	-	&	-	\\
4923.927	&		&	2.891	&	-1.320	&	-	&	-	&	-	&	-	&	86.1(5.29)	&	-	&	79.2(5.25)	\\
5197.577	&		&	3.231	&	-2.100	&	-	&	-	&	62.8(5.31)	&	36.7(5.16)	&	38.5(5.28)	&	-	&	-	\\
5991.376	&		&	3.153	&	-3.557	&	31.4(6.29)	&	-	&	-	&	-	&	-	&	-	&	-	\\

\hline
\end{tabular}}
The numbers in the  parenthesis in columns 5-11 give the derived abundances from the respective line.
log gf values are taken from  kurucz atomic line list (\url{https://www.cfa.harvard.edu}).
\end{table*}
}

Only those lines with excitation potential from 0 - 5 eV and equivalent widths from 20 - 180 m{\rm \AA} are considered for the analysis. We have made use of MOOG (Sneden 1970, updated version 2013) for the analysis under the assumption of local thermodynamic equilibrium. We have used alpha normal model atmospheres ([$\alpha$/Fe] = 0) for our analysis and the model atmospheres are adopted from the Kurucz grid of model atmospheres with no convective overshooting (\url{http://cfaku5.cfa.hardvard.edu/}). Solar abundances are taken from \cite{asplund.2009ARA&A..47..481A}. The derived atmospheric parameters of our programme stars along with the literature values are presented in Table \ref{table5}.

{\footnotesize
\begin{table*}
\caption{\bf Derived atmospheric parameters of our programme stars and literature values. }
\begin{tabular}{lccccccc}
\hline
Star            &T$_{eff}$  & log g       &$\zeta$            & [Fe I/H]         &[Fe II/H] & [Fe/H] & Reference\\
                &(K)        & cgs         &(km s$^{-1}$)      &                  &          &        &\\
                &($\pm$100) & ($\pm$0.2)  & ($\pm$0.2)        &                  &          &        & \\
\hline
HE 0110$-$0406  &	4670	&	1.00	          &	1.92	  &	 $-1.31$$\pm$0.09 (27)	&	 $-1.29$$\pm$0.12 (2)	& $-1.30$ & 1\\
HE 1425$-$2052 	&	4300	&	1.50          	&	2.70 	  &	 $-2.10$$\pm$0.20 (28)	&	$-2.09$$\pm$0.16 (2)	& $-2.09$ & 1 \\
HE 1428$-$1950 	&	4500	&	0.50	          &	2.75	  &	 $-2.11$$\pm$0.16 (27)	&	$-2.15$$\pm$0.16 (3)	& $-2.11$ & 1\\
                & 4562  & 3.75            &  -      & -                 &  -                & $-2.07$ & 2 \\
                & 4531  & 0.85            &  -      & -                 &  -                & $-2.07$ & 3 \\
HE 1429$-$0551 	&	4940	&	1.50	          &	1.54	  &	 $-2.47$$\pm$0.15 (31)	&	$-2.44$$\pm$0.13 (2) 	& $-2.45$ & 1 \\
                & 4700  & 1.50            & 2.00    &  $-2.47$$\pm$0.20 (17) & $-2.48$$\pm$0.16 (3)  & $-2.47$ & 4 \\
                & 4832  & 1.14            & 2.01    &   -               &  -                & $-2.70$ & 5 \\
HE 1447+0102 	  &	5220	&	1.90	          &	1.36	  &	 $-2.29$$\pm$0.12 (20)&	$-2.21$$\pm$0.00 (2) 	& $-2.25$ & 1 \\
                & 5100  & 1.70            & 1.80    & $-2.47$$\pm$0.27  (12) & $-2.45$$\pm$0.14 (3) & $-2.46$ & 4 \\
HE 1523$-$1155 	&	4780	&	1.60	          &	1.52	  &	 $-2.43$$\pm$0.11  (30)&	$-2.44$  (1)        	& $-2.42$ & 1 \\
                & 4800  & 1.60            & 1.80    & $-2.15$$\pm$0.19  (16) & $-2.17$$\pm$0.14  (3)  & $-2.16$ & 4 \\
HE 1528$-$0409 	&	5250	&	2.00	          &	1.29	  &	 $-2.37$$\pm$0.13  (19)	&	$-2.34$$\pm$0.12 (2) 	& $-2.35$ & 1\\
                & 5000  & 1.80            & 1.80    & $-2.61$$\pm$0.19 (15)  & $-2.59$$\pm$0.14 (1) & $-2.60$ & 4 \\
\hline
\end{tabular}

1. Our work, 2. \cite{placco.2011AJ....142..188P},  3. \cite{kennedy.2011AJ....141..102K}, 4. \cite{aoki.2007ApJ...655..492A}, 5. \cite{karinkuzhi.2021A&A...645A..61K}\\ \textit{Note} : The number of lines used for analysis is given inside the parenthesis.
\label{table5}
\end{table*}
} 

\section{Abundance analysis}\label{sec:abund-analysis}
Abundances of various elements are determined using the equivalent width measurements as well as using the spectrum synthesis calculation of clean and unblended lines. Lines are identified by over plotting the Arcturus spectrum on the spectra of the programme stars. A master line list is prepared using the measured equivalent widths of the lines and the other line information such as excitation potential and log\,{gf} values that are taken from the Kurucz database. Abundances of light elements, C and N, odd-Z element Na, $\alpha$ elements such as Mg, Ca, Sc and Ti and Fe-peak elements such as Cr, Mn, Co and Ni are estimated whenever possible. Abundances of neutron-capture elements Sr, Y, Ba, La, Ce, Pr, Nd, Sm and Eu are also determined whenever the lines are available. For abundance determination of the elements Sc, V, Mn, Ba, La and Eu, we have also used spectrum synthesis calculation considering their hyperfine structures taken from various sources. Estimated abundances are presented in Tables \ref{table6} - \ref{table8}.
\subsection{Carbon, Nitrogen, and Oxygen}
\begin{figure}
\centering
\includegraphics[width=12cm,height=11cm]{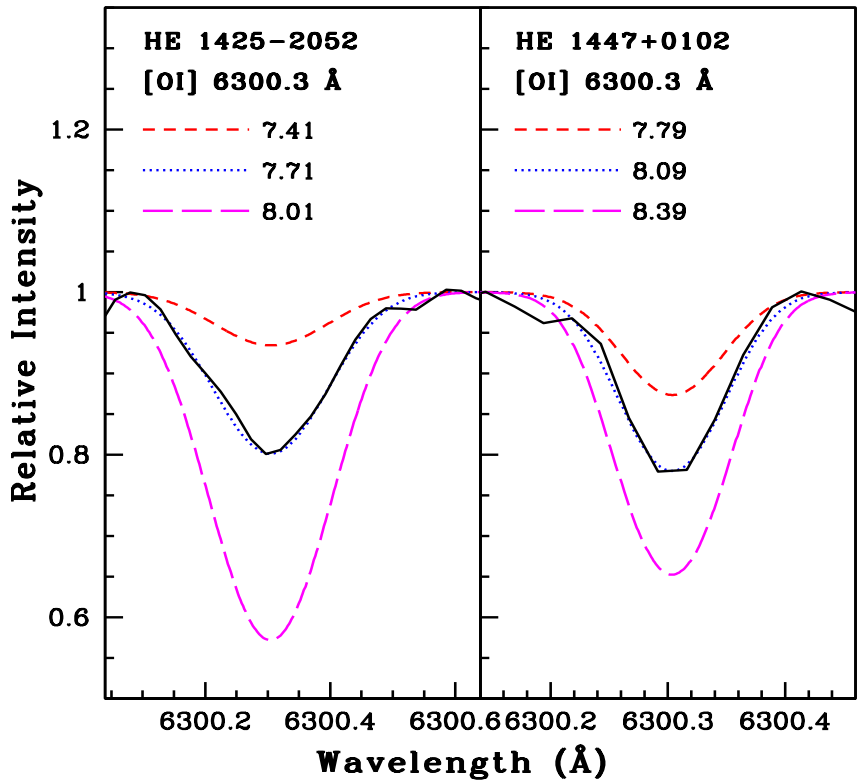}
\caption{ Synthesis of [OI] line around 6300 {\rm \AA}. Dotted line represents synthesized spectra and the solid line indicates the observed spectra. Short dashed line represents the synthetic spectra corresponding to $\Delta$[O/Fe] = -0.2 and long dashed line corresponds to $\Delta$[O/Fe] = +0.2}
\label{fig3}
\end{figure}

The abundance of oxygen is determined from the spectrum synthesis of the oxygen forbidden line [OI] 6300.3 {\rm \AA} (Figure \ref{fig3}). We could determine O in all the programme stars except HE 0110$-$0406. In HE 0110$-$0406, [OI] 6300.3 {\rm \AA} and O I 6363.7 {\rm \AA} lines are found to be blended and not suitable for abundance determination. O is enhanced in all the stars with [O/Fe] ranging from 1.12 to 1.88 , except for HE 1523$-$1155 and HE 1528$-$0409. The line list used for the synthesis of O is taken from Kurucz database.

\par
\begin{figure}
\centering
\includegraphics[width=12cm,height=11cm]{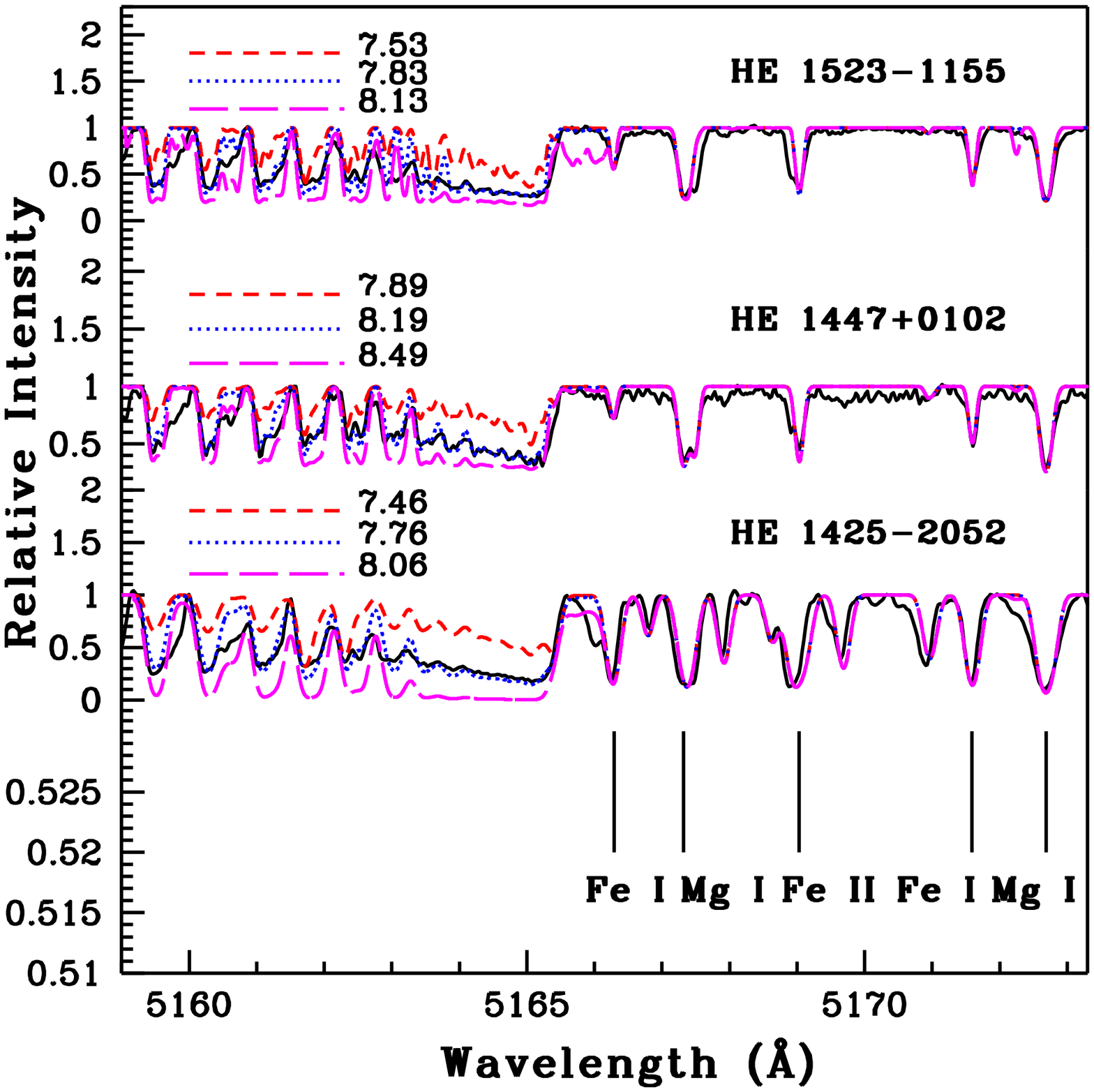}
\caption{Synthesis of C$_{2}$ band around 5165 {\rm \AA}. Dotted line represents synthesized spectra and the solid line indicates the 
observed spectra. Short dashed line represents the synthetic spectra corresponding to $\Delta$ [C/Fe] = -0.2 and long dashed line corresponds to $\Delta$[C/Fe] = +0.2}
\label{fig4}
\end{figure}

\begin{figure}
\centering
\includegraphics[width=12cm,height=11cm]{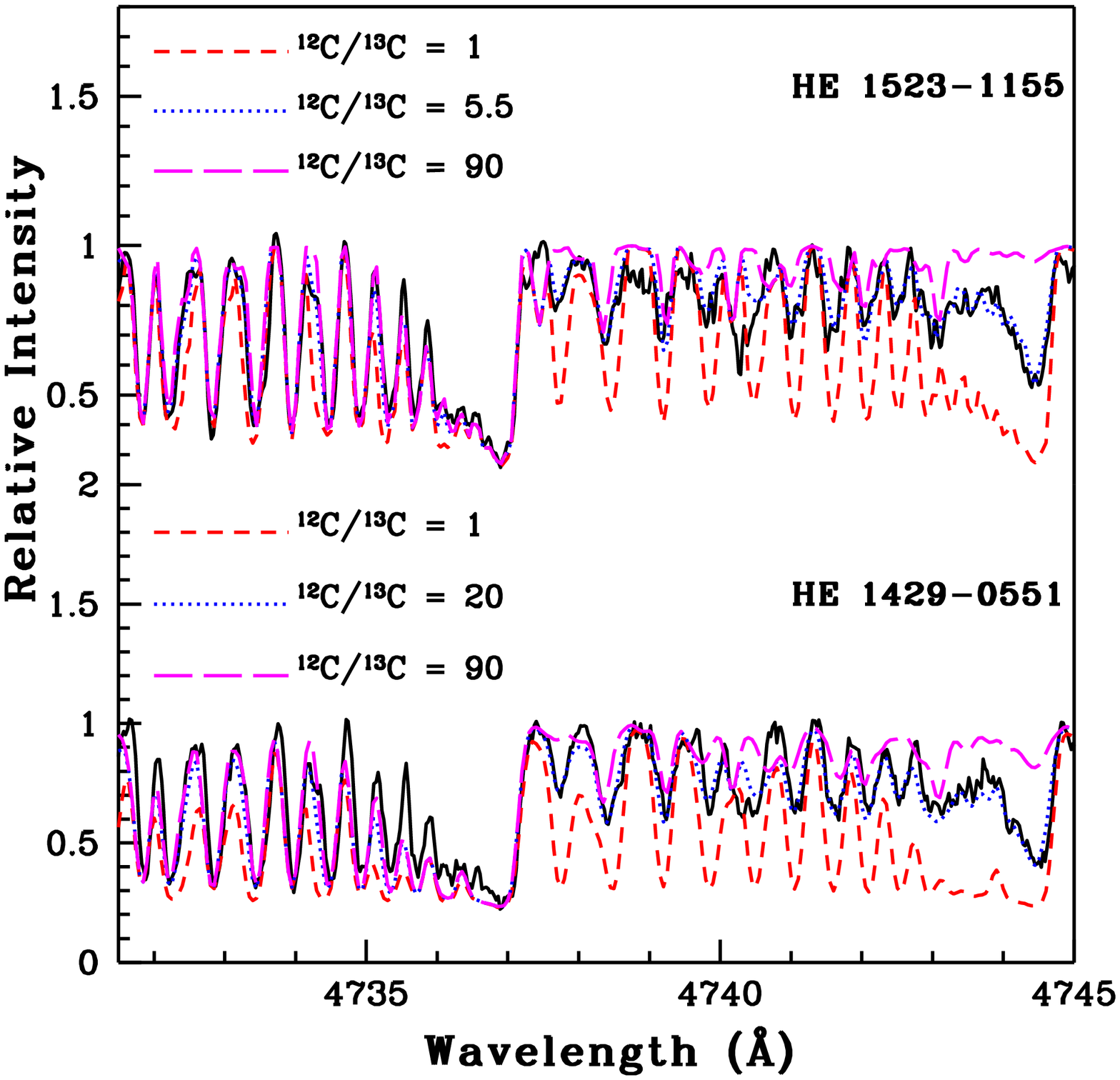}
\caption{Spectral synthesis fits of the C$_{2}$ features around 4740 {\rm \AA}. Solid line indicates the observed spectra. Short and long dashed lines are shown to illustrate the sensitivity of the line strengths to the isotopic carbon abundance ratios}
\label{fig5}
\end{figure}

 The abundance of carbon is determined using the spectrum synthesis calculation of the C$_{2}$ band at 5165 {\rm \AA} (Figure \ref{fig4}).  We could also determine the C abundance using the spectrum  synthesis calculation of CH band around 4315 {\rm \AA} only for HE 1428$-$1950 and HE 1425$-$2052. For the remaining stars CH band is found to be saturated except HE 0110$-$0406, and HE 1528$-$0409 for which this region is noisy and not suitable for abundance determination. Since we could determine the abundance of C from the synthesis of C$_{2}$ band at 5165 {\rm \AA} for all the stars, we have used this value throughout our analysis. C is enhanced in all the programme stars with [C/Fe] ranging from 0.73 to 2.29. We could estimate the carbon isotopic ratio in all the programme stars. Carbon isotopic ratio is determined from the spectrum synthesis calculation of the C$_{2}$ swan system around 4740{\rm \AA} (Figure \ref{fig5}). The estimated values are presented in Table \ref{table9}.

\par
\begin{figure}
\centering
\includegraphics[width=12cm,height=11cm]{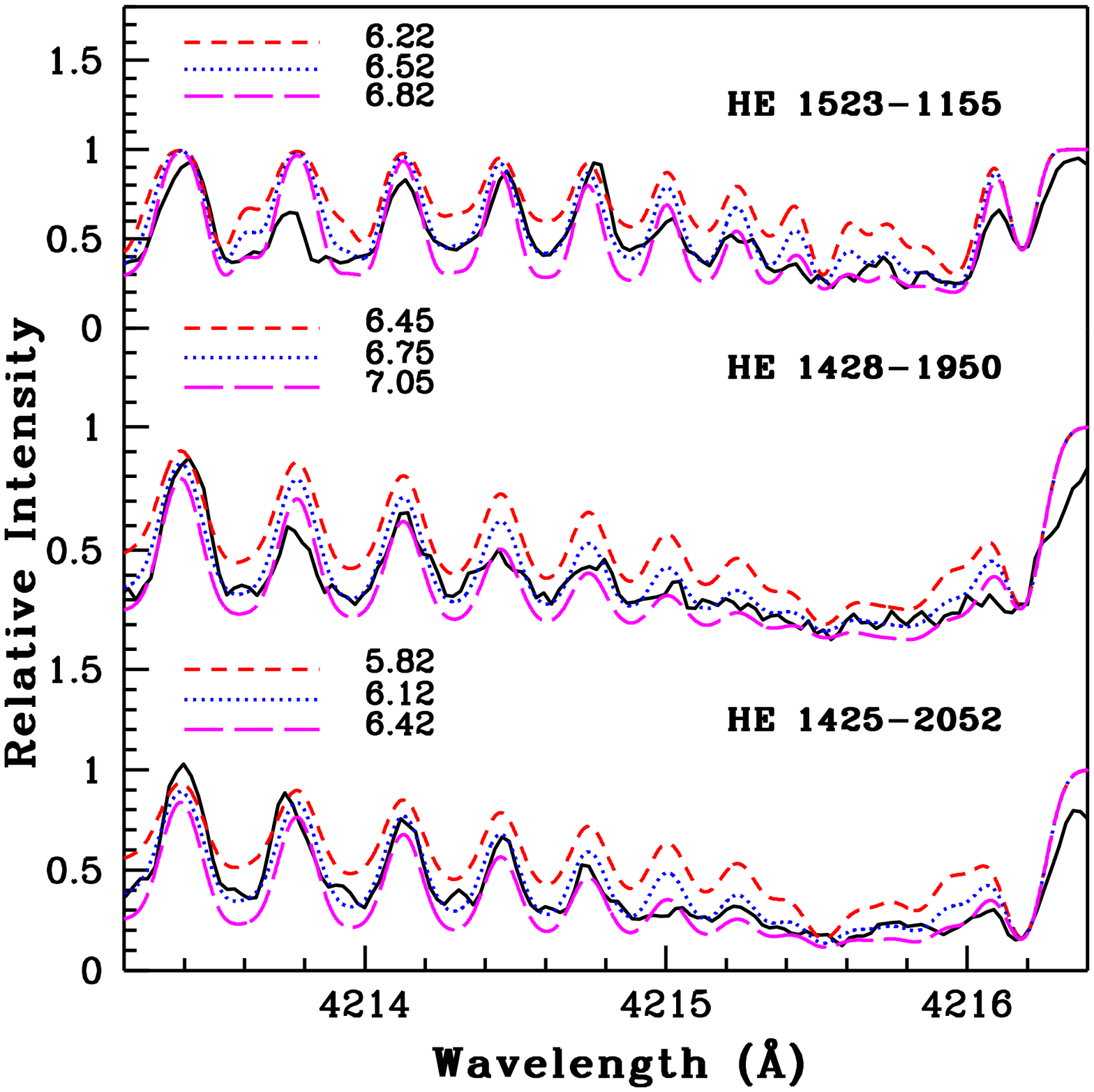}
\caption{ Synthesis of CN band around 4215 {\rm \AA}. Dotted line represents synthesized spectra and the solid line indicates the observed spectra. Short dashed line represents the synthetic spectra corresponding to $\Delta$[N/Fe] = -0.2 and long dashed line corresponds to $\Delta$[N/Fe] = +0.2}
\label{fig6}
\end{figure}

 Nitrogen abundance is determined from the spectrum synthesis calculation of CN band at 4215{\rm \AA} (Figure \ref{fig6}).  N is enhanced in all the programme stars except HE 1425$-$2052 with [N/Fe] $\sim$ 0.39. The molecular line lists for C and N are adopted from \cite{sneden.2014ApJS..214...26S} and \cite{ram.2014ApJS..211....5R}.

\subsection{Na, Mg, Ca, Sc, Ti, and V}
Abundance of sodium is determined from the equivalent width measurements of three Na I lines (Table \ref{tableA1}).
HE 1428$-$1950 shows enhancement of Na with [Na/Fe] $\sim$ 1.59. In the remaining objects, Na is moderately enhanced with [Na/Fe] ranging from 0.25 to 0.86, except HE 1528$-$0409 for which [Na/Fe] $\sim$ 0.25. Abundance of magnesium is estimated using the equivalent width measurement of three Mg I lines (Table \ref{tableA1}). The estimated values of Mg abundance lie in the range [Mg/Fe] $\sim$ 0.27 to 0.96.
 
\par We have used the equivalent width measurements of 11 Calcium I lines (Table \ref{tableA1}) to estimate the abundance of  Ca in the programme stars. Ca is near solar in HE 1429$-$0551 and HE 0110$-$0406. In the remaining objects, [Ca/Fe] values range from 0.23 to 0.47. The abundance of Scandium is determined using the spectrum synthesis calculation of the lines Sc II 4415.56, 4431.350 and 6245.63 using the hyperfine structures  from \cite{prochaska.2000ApJ...537L..57P}. Sc abundance is found to be near solar in all the programme stars except for HE 1425$-$2052 and HE 1523$-$1155 for which [Sc/Fe] values are 0.44 and 0.35 respectively.

\par The abundance of titanium  is determined using the equivalent width measurements of Ti I and Ti II lines (Table \ref{tableA1}). 
Ti  is found to be under abundant in HE 1447+0102 with [Ti/Fe] $\sim$ $-0.28$. In the remaining objects, [Ti/Fe] values are in the range 0.12 to 0.43. The abundance of Vanadium is determined from the spectrum synthesis calculation of the lines V I 4864.73 {\rm \AA} and 5727.05 {\rm \AA} using the hyperfine structures  from \cite{prochaska.2000ApJ...537L..57P}. HE 1428$-$1950 and HE 1528$-$0409 show enhancement of V with [V/Fe] $\sim$ 0.90. While the abundance of V is moderately enhanced in HE 1425$-$2052 and HE 1447+0102, V is marginally enhanced in HE 1429$-$0551 and HE 1523$-$1155. V is found to be under abundant in HE 0110$-$0406.  

\subsection{Cr, Mn, Co, Ni, and Zn}

Abundance of cromium  is estimated using the equivalent width measurements of a number of Cr I and Cr II lines (Table \ref{tableA1}).
While Cr is enhanced in HE 1428$-$1950 with [Cr/Fe] $\sim$ 1.29, it is found to be under abundant in HE 1425$-$2052 with [Cr/Fe] $\sim$ $-0.26$. HE 1523$-$1155 shows moderate enhancement of Cr and near solar values for HE 1429$-$0551 and HE 0110$-$0406. We could not measure Cr in HE 1447+0102 as the lines are found to be blended and not suitable for abundance determination. The abundance of Manganese is determined using the spectrum synthesis calculation of the lines Mn I 4470.14 {\rm \AA} and 6013.51 {\rm \AA} considering the hyperfine structure from \cite{prochaska.2000ApJ...537L..57P}. [Mn/Fe] values in the programme stars range from $-0.25$ to 0.46.

\par The abundance of cobalt could be determined only in HE 0110$-$0406, HE 1425$-$2052 and HE 1428$-$1950. The Co I lines are found to be blended in the remaining objects. Co is enhanced in HE 1428$-$1950 with [Co/Fe] $\sim$ 0.93 and near solar in HE 0110$-$0406 and HE 1425$-$2052. The abundance of  Nickel could be determined in all the programme stars except for HE 1528$-$0409. In all the stars, Ni is found to be slightly enhanced with [Ni/Fe] in the range 0.26 to 0.45. The abundance of Zinc could be determined only in HE 0110$-$0406, HE 1425$-$2052, HE 1428$-$1950 and HE 1523$-$1155 with [Zn/Fe] in the range $-0.41$ to 0.50. In the remaining stars, the Zn I lines are found to be highly blended and could not be used for abundance 
calculation.

\subsection{Sr, Y, and Zr}

The abundance of strontium  is determined from the spectrum synthesis calculation of the line Sr I 4607.33 {\rm \AA}. Sr is enhanced in all the programme stars. We could not estimate the abundance of Sr in HE 1429$-$0551 and HE 1528$-$0409 as this region is very noisy and not suitable for abundance calculation. 

\par We have used the equivalent width measurements of three Yttrium II lines (Table \ref{tableA1}) for the determination of Y abundance in the programme stars. While Y is enhanced in HE 0110$-$0406, HE 1425$-$2052 and HE 1428$-$1950 with [Y/Fe] $\sim$ 1.28, 1.37, and 1.25 respectively, it is only moderately enhanced in HE 1523$-$1155 with [Y/Fe] $\sim$ 0.68. In the remaining stars, the abundance of Y could not be determined as the lines are blended. We could not determine the abundance of zirconium in any of our programme stars as the lines due to Zr are either blended or the region is found to be noisy and not suitable for abundance determination.
 
\subsection{Ba, La, Ce, Pr, Nd, Sm, and Eu}

Abundance of barium is determined from the spectrum synthesis calculations of the lines Ba II 5853.66 and 6141.71 {\rm \AA} using the hyperfine structures of Ba from \cite{McWilliam.1998AJ....115.1640M}. We could not estimate the Ba abundance in HE 1428$-$1950 as the lines Ba II 5853.66 and 6141.71{\rm \AA} are saturated. We have also checked other lines due to Ba such as 4554.03 and 4934.08 {\rm \AA} in this object; these lines are found to be highly blended and not suitable for abundance calculation. Ba is enhanced in all the remaining programme stars with [Ba/Fe] ranging from 1.61 to 3.21. In HE 1425$-$2052, Ba is found to be highly enhanced with [Ba/Fe] $\sim$ 3.21. We have used the spectrum synthesis calculation of the line Lanthanum II 4921.78 using the hyperfine structure  from \cite{jonsel.2006A&A...451..651J}. La is enhanced in all the programme stars with [La/Fe] ranging from $\sim$1.04 to 2.49. The abundance of Cerium is estimated from the equivalent width measurements of ten Ce II lines (Table \ref{tableA1}).
All the programme stars exhibit enhancement of Ce with [Ce/Fe] ranging from 1.14 to 2.04. 

\par We could determine the abundance of praseodymium only in HE 0110$-$0406, HE 1429$-$0551, HE 1447+0102 and HE 1528$-$0409. In all these stars Pr is found to be enhanced with [Pr/Fe] ranging from 1.09 to 2.05. The abundance of neodymium could be measured in all the programme stars and all of them exhibit enhancement of Nd with [Nd/Fe] in the range 1.03 to 2.34. We could estimate the abundance of samarium in all the programme stars with [Sm/Fe] in the  range  1.15 to 2.66. We have used the spectrum synthesis calculation of the lines Europium II 4129.71, 6437.63 and 6645.11 {\rm \AA} to determine the abundance of Eu in the programme stars. The hyperfine structures of Eu are taken from \cite{worley.2013A&A...553A..47W}. Eu is enhanced in all the programme stars except HE 0110$-$0406 and HE 1528$-$0409. 
\par We have checked for the lines due to the heavy elements such as Ru I 4869.15 {\rm \AA}, several Nb lines Nb I 4152.50, 5642.09, 5729.18, and 5983.2 {\rm \AA}, Mo I lines at 4904.39, and 5196.04 Å, and Dy I 4923.16 {\rm \AA}. In our programme stars, all these lines are found to either absent or very weak and blended with equivalent widths less than 10m{\rm \AA}. We have also checked the line Pb I 4057.7{\rm \AA} in the spectrum of HE 0110-0406 and this region is found to be noisy. 
\par We have estimated the values of [ls/Fe], [hs/Fe], [hs/ls] where ls represents light s-process elements (Sr, Y, Zr) and hs represents the heavy s-process elements (Ba, La, Ce, Nd). The estimated values of [hs/ls] and the carbon isotopic ratios are presented in Table \ref{table9}. Table \ref{table10} presents a comparison of the abundances of our programme stars with the literature values.

{\footnotesize
\begin{table*}
\caption{\bf Elemental abundances in HE 0110$-$0406 }
\begin{tabular}{lccccccccc}
\hline                      
 & Z & solar log$\epsilon^{\ast}$ & log$\epsilon$ & [X/H] & [X/Fe] \\
\hline
C (C$_{2}$, 5165 {\rm \AA}) & 6 & 8.43 & 7.85$\pm$0.20(syn)   & $-0.58$ & 0.73 \\
N (CN, 4215 {\rm \AA}) & 7 & 7.83 & 7.15$\pm$0.20(syn)   & $-0.68$ & 0.63 \\
Na I & 11 & 6.24 & 5.53(1) & $-0.71$ & 0.60  \\
Mg I & 12 & 7.60 & 6.56(1) & $-1.04$ & 0.27  \\
Ca I & 20 & 6.34 & 5.02$\pm$0.13(4) & $-1.32$ & $-0.01$\\
Sc II& 21 & 3.15 & 1.70$\pm$0.20(1,syn)& $-1.45$    & $-0.14$ \\
Ti I & 22 & 4.95 & 3.76$\pm$0.14(4) & $-1.19$    & 0.12 \\
V I  & 23 & 3.93 & 2.17$\pm$0.20(1,syn) & $-1.76$    & $-0.45$ \\
Cr I & 24 & 5.64 & 4.29$\pm$0.20(3) & $-1.35$    & $-0.04$ \\
Mn I & 25 & 5.43 & 4.30$\pm$0.20(1,syn)         & $-1.13$ & 0.18\\
Fe I & 26 & 7.50 & 6.19$\pm$0.09(27)& $-1.31$    & -\\
Fe II& 26 & 7.50 & 6.21$\pm$0.12(2)& $-1.29$    & -\\
Co I & 27 & 4.99 & 3.73$\pm$0.04(3)           & $-1.26$    & 0.05 \\
Ni I & 28 & 6.22 & 5.36(1)& $-0.86$ & 0.45\\
Zn I & 30 & 4.56 & 2.84(1)          & $-1.72$ & $-0.41$ \\
Sr I & 38 & 2.87 & 2.34$\pm$0.20(1,syn)          & $-0.53$    & 0.78  \\
Y II & 39 & 2.21 & 2.20(1) & $-0.01$    & 1.28\\
Ba II& 56 & 2.18 & 2.50$\pm$0.20(1,syn)           & 0.32    & 1.61\\
La II& 57 & 1.10 & 1.07$\pm$0.20(1,syn)          & $-0.03$    & 1.26 \\
Ce II& 58 & 1.58 & 1.49$\pm$0.21(3) & $-0.09$    & 1.20 \\
Pr II& 59 & 0.72 & 0.82$\pm$0.09(2)           & 0.10    & 1.39 \\
Nd II& 60 & 1.42 & 1.49$\pm$ 0.18(4) & 0.07    & 1.36\\
Sm II& 62 & 0.96 & 0.90$\pm$0.14(4)           & $-0.06$    & 1.23\\ 
Eu II& 63 & 0.52 & $-0.25$$\pm$0.20{\bf (1,syn)}         & $-0.77$    & 0.52 \\
\hline
\end{tabular}

$^\ast$  \cite{asplund.2009ARA&A..47..481A}, The number inside the parenthesis shows the number 
of lines used for the abundance determination.
\label{table6}
\end{table*}

} 
 
{\footnotesize
\begin{table*}
\caption{\bf Elemental abundances in HE 1425$-$2052, HE 1428$-$1950 and HE 1429$-$0551}
\resizebox{\textwidth}{!}{\begin{tabular}{|ccc|ccc|ccc|ccc|}
\hline
  &      &       & \multicolumn{3}{c}{HE 1425$-$2052} & \multicolumn{3}{c}{HE 1428$-$1950} & \multicolumn{3}{c}{HE 1429$-$0551} \\
\hline
  & Z & Solar log$\epsilon^{a}$ & log$\epsilon$& [X/H]& [X/Fe]&log$\epsilon$& [X/H]    &[X/Fe]& log$\epsilon$  & [X/H] & [X/Fe] \\ \hline
C (C$_{2}$, 5165 {\rm \AA})    & 6  & 8.43  & 7.76$\pm$0.20(syn)         & $-0.67$ & 1.43 & 7.92$\pm$0.20(syn)         & $-0.51$ & 1.60 & 8.25$\pm$0.20(syn)         & $-0.18$ & 2.29\\
C (CH, 4315 {\rm \AA}) & 6 &  8.43 & 7.72$\pm$0.20(syn)   & $-0.71$& 1.39 & 7.95$\pm$0.20(syn) & $-0.48$ & 1.63 &  - &  - &  - \\
N (CN, 4215 {\rm \AA})    & 7  & 7.83  & 6.12$\pm$0.20(syn)         & $-$1.71 & 0.39 & 6.75$\pm$0.20(syn)         & $-1.08$ & 1.03 & 6.44$\pm$0.20(syn)         & $-1.39$ & 1.08 \\
O     & 8  & 8.69  & 7.71$\pm$0.20(syn)         & $-0.98$ & 1.12 & 7.84$\pm$0.20(syn)         & $-0.85$ & 1.26 & 8.10$\pm$0.20(syn)         & $-0.59$ & 1.88 \\
Na I	& 11 &	6.24 & 4.75(1)           & $-$1.49 & 0.61 & 5.72$\pm$0.20(1,syn)         & $-0.52$ & 1.59 & 4.63(1)           & $-1.61$ & 0.86  \\
Mg I	& 12 &	7.60 & 5.83(1)           & $-$1.77 & 0.33 & 5.92$\pm$0.20(1,syn)         & $-1.68$ & 0.43 & 5.55(1)           & $-2.05$ & 0.42\\
Ca I	& 20 &	6.34 & 4.59$\pm$0.20 (6) & $-$1.75 & 0.35 & 4.70$\pm$0.20(2)  & $-1.64$ & 0.47 & 3.92$\pm$0.10(3)  & $-2.42$ & 0.05\\
Sc II	& 21 &	3.15 & 1.50$\pm$0.20(1,syn)         & $-$1.65 & 0.44 &1.09$\pm$0.20(1,syn)          & $-2.06$ & 0.09 & 0.62$\pm$0.20(1,syn)         & $-2.53$ & $-0.09$ \\
Ti I	& 22 &	4.95 & 3.20$\pm$0.13 (2) & $-$1.75 & 0.35 & -                 & -        &-     & 2.75(1)     & $-2.20$ & 0.27\\
Ti II	& 22 &	4.95 & 3.15$\pm$0.20(2)  & $-$1.80 & 0.29 & 3.14$\pm$0.20(2)  & $-1.81$  & 0.34 & 2.79$\pm$0.18(3)  & $-2.16$ & 0.28\\
V I	  & 23 &	3.93 & 2.33$\pm$0.20(1,syn)         & $-$1.60 &0.50  & 2.73$\pm$0.20(1,syn)         & $-1.20$  & 0.91 & 1.81$\pm$0.20(1,syn)    &    $-2.12$    &  0.35  \\
Cr I	& 24 &	5.64 & 3.28(1)           & $-2.36$ &$-0.26$& 4.81$\pm$0.16(2) & $-0.83$  & 1.28 & 3.18(1)    & $-2.46$ & 0.01 \\
Cr II	& 24 &	5.64 & -                 &   -     & -    &  4.79(1)          & $-0.85$  & 1.30 & - & - & -\\
Mn I	& 25 &	5.43 & 3.08$\pm$0.20(1,syn)         & $-$2.35 &$-0.25$& 3.66$\pm$0.20(1,syn)        & $-1.77$  &0.34  &3.42$\pm$0.20(1,syn)         & $-2.01$ & 0.46 \\
Fe I	& 26 &	7.50 & 5.40$\pm$0.20(28) & $-$2.10 & -    & 5.39$\pm$0.16(27) & $-2.11$ & -    & 5.03$\pm$0.15(31) & $-2.47$ & - \\
Fe II	& 26 &	7.50 & 5.41$\pm$0.16(2)  & $-$2.09 & -    & 5.35$\pm$0.16(3)  & $-2.15$ & -    & 5.06$\pm$0.13(2)  & $-2.44$ & - \\
Co I	& 27 &	4.99 & 2.97(1)           & $-$2.02 & 0.08 & 3.81(1)           & $-1.18$ & 0.93 & -                 & -       & - \\
Ni I	& 28 &	6.22 & 4.52$\pm$0.16(2)  & $-$1.70 & 0.40 & 4.38$\pm$0.05(2)  & $-1.84$ & 0.27 &4.01(1)      & $-2.21$ & 0.26 \\
Zn I	& 30 &	4.56 & 2.88(1)           &$-$1.68  & 0.42 & 2.95(1)           & $-1.61$ & 0.50 & -                 & -       & -\\ 
Sr I	& 38 &	2.87 & 2.31$\pm$0.20(1,syn)         &$-$0.56  & 1.53 & 1.80$\pm$0.20(1,syn)         & $-1.07$ & 1.04 & -                 & -       & - \\
Y II	& 39 &	2.21 & 1.49$\pm$0.04(2)  & $-$0.72 & 1.37 & 1.30(1)         & $-0.91$ & 1.25 & -                 & -       & - \\
Ba II	& 56 &	2.18 & 3.30$\pm$0.20(1,syn)         & 1.12    & 3.21 & -                 & -       & -    & 1.62$\pm$0.20(1,syn)         & $-0.56$ & 1.88 \\
La II	& 57 &	1.10 & 1.49$\pm$0.20(1,syn)         & 0.39    & 2.49 &$-0.01$$\pm$0.20{\bf (1,syn)}       & $-1.11$ & 1.04 & $-0.20$$\pm$0.20(1,syn)      & $-1.30$ & 1.14 \\
Ce II	& 58 &	1.58 & 1.70(1)           & $-0.14$ & 1.95 & 0.57$\pm$0.17(3)  & $-1.01$ & 1.14 & 1.04$\pm$0.20(4)     & $-0.54$ & 1.90 \\
Pr II & 59 & 0.72  & -     & -       & -    & -                 & -       & -    & $-0.63$(1)    & $-1.35$ & 1.09   \\            
Nd II	& 60 &	1.42 & 1.67$\pm$0.20(6)  & 0.25    & 2.34 & 0.30$\pm$0.06(3)  & $-1.12$ & 1.03 & 0.72$\pm$0.15(7)     & $-0.70$ & 1.74 \\
Sm II	& 62 &	0.96 & 1.53$\pm$0.20(6)  & 0.57    & 2.66 & $-0.04$(1)        & $1.00$  & 1.15 & 0.29(1)              & $-0.67$ & 1.77\\            
Eu II & 63 &  0.52 & 0.41$\pm$0.20(1,syn)         & $-0.11$ & 1.98 & $-0.56$$\pm$0.20{\bf (1,syn)}      & $-1.08$ & 1.07 & $-0.67$$\pm$0.20(1,syn)      & $-1.19$ & 1.25 \\
\hline
\end{tabular}
}

$^a$  \cite{asplund.2009ARA&A..47..481A}, The number inside the  parenthesis shows the 
number of lines used for the abundance determination. 
\label{table7}
\end{table*}
} 

{\footnotesize
\begin{table*}
\caption{\bf Elemental abundances in HE 1447+0102, HE 1523$-$1155 and HE 1528$-$0409 }
\resizebox{\textwidth}{!}{\begin{tabular}{|ccc|ccc|ccc|ccc|}
\hline
  &      &       & \multicolumn{3}{c}{HE 1447+0102} & \multicolumn{3}{c}{HE 1523$-$1155} & \multicolumn{3}{c}{HE 1528$-$0409} \\
\hline
  & Z & Solar log$\epsilon^{a}$ & log$\epsilon$& [X/H]& [X/Fe]&log$\epsilon$& [X/H]    &[X/Fe]& log$\epsilon$  & [X/H] & [X/Fe] \\ \hline
C (C$_{2}$, 5165 {\rm \AA})    & 6  & 8.43  & 8.19$\pm$0.20(syn)         & $-0.24$ & 2.05 & 7.83$\pm$0.20(syn)          & $-0.60$ & 1.83 & 8.13$\pm$0.20(syn)          & $-0.30$    & 2.07\\
N (CN, 4215 {\rm \AA})    & 7  & 7.83  & 7.13$\pm$0.20(syn)         & $-0.70$ & 1.59 & 6.52$\pm$0.20(syn)          & $-1.31$ & 1.12 & 7.13$\pm$0.20(syn)         & $-0.7$     & 1.67 \\
O     & 8  & 8.69  & 8.09$\pm$0.20(syn)         & $-0.60$ & 1.69  & 7.08$\pm$0.20(syn)         & $-1.61$ & 0.82 & 6.91$\pm$0.20(syn)         & $-1.78$    & 0.59 \\
Na I	& 11 &	6.24 &4.60$\pm$0.05(2)   & $-1.64$ & 0.65  & 4.53(1)           & $-1.71$ & 0.72 & 4.12$\pm$0.08(2)  & $-2.12$    & 0.25   \\
Mg I	& 12 &	7.60 &  5.88$\pm$0.19(2) & $-1.72$ & 0.57  & 6.13(1)           & $-1.47$ & 0.96& 5.80(1)     & $-1.80$    & 0.57\\
Ca I	& 20 &	6.34 & 4.28$\pm$0.14(5) & $-2.06$ & 0.23  & 4.26$\pm$0.10(5)   & $-2.08$ & 0.35 & 4.28$\pm$0.12(5)       & $-2.06$    & 0.31\\
Sc II	& 21 &	3.15 & 1.18$\pm$0.20(1,syn)       & $-1.96$ & 0.24  & 1.05$\pm$0.20(1,syn)         & $-2.10$ & 0.35   & 0.90$\pm$0.20(1,syn)           & $-2.25$    & 0.09 \\
Ti I	& 22 &	4.95 & -                & -       &-      & 2.85(1)           & $-2.10$ & 0.33   & -                   & -          & -\\
Ti II	& 22 &	4.95 & 2.46$\pm$0.13(4)& $-2.49$& $-0.28$ & 2.93$\pm$0.10(4)  & $-2.02$ & 0.43  & 2.81$\pm$0.13(3)      & $-2.14$    & 0.20\\
V I	  & 23 &	3.93 & 2.27$\pm$0.20(1,syn)           & $-1.66$ & 0.63 &1.81$\pm$0.20(1,syn)             & $-2.12$ & 0.31   & 2.46$\pm$0.20(1,syn)         & $-1.47$    & 0.90   \\
Cr I	& 24 &	5.64 & -                 &   -     & -    & 3.73$\pm$0.16(2)   & $-1.91$ & 0.52 & -                  & -          & - \\
Cr II	& 24 &	5.64 & 3.72              &  $-1.92$ & 0.29 & 3.64(1)           & $-2.00$ & 0.45   & -                  & -          & - \\
Mn I	& 25 &	5.43 & 3.60$\pm$0.20(1,syn)         & $-1.83$ &0.46   & -                & -       &-      &-                  & -          & -\\
Fe I	& 26 &	7.50 & 5.21$\pm$0.12(20) & $-2.29$ & -    & 5.07$\pm$0.11(30)& $-2.43$ & -    & 5.13$\pm$0.13(19)  & $-2.37$ & - \\
Fe II	& 26 &	7.50 & 5.29$\pm$0.0(2)   & $-2.21$ & -     & 5.06(1)         & $-2.44$ & -    & 5.16$\pm$0.12(2)  & $-2.34$ & - \\
Ni I	& 28 &	6.22 & 4.37(1)         & $-1.85$ &0.44   & 4.20$\pm$0.12(3)        & $-2.02$ &0.41&-                  & -       & - \\
Zn I	& 30 &	4.56 &  -                &  -      &  -  &  2.54(1)           &  $-2.02$ & 0.41   & -                   & -       & - \\ 
Sr I	& 38 &	2.87 & 2.15$\pm$0.20(1,syn)         & $-0.72$ & 1.57 & 1.37$\pm$0.20(1,syn)         & $-1.50$ & 0.93 & -                  & -       & - \\
Y II	& 39 &	2.21 & -                 & -       & -  & 0.44(1) & $-1.77$ & 0.68 & -                    & -       & - \\
Ba II	& 56 &	2.18 & 1.90$\pm$0.20(1,syn)         & $-0.28$ & 1.93 & 1.57$\pm$0.20(1,syn)         & $-0.61$ & 1.83 & 1.55$\pm$0.20(1,syn)          & $-0.63$ & 1.71 \\
La II	& 57 &	1.10 & 1.01$\pm$0.20(1,syn)         & $-0.09$ & 2.12    & 0.24$\pm$0.20$\pm$0.20(1,syn)      & $-0.86$ & 1.59 & 0.09$\pm$0.20(1,syn)       & $-1.01$ & 1.33 \\
Ce II	& 58 &	1.58 & 1.41$\pm$0.12(3)     & $-0.17$ & 2.04 & 0.70$\pm$0.14(4) & $-0.88$ & 1.57 & 1.13$\pm$0.20(2)    & $-0.45$ & 1.89 \\
Pr II & 59 & 0.72  & 0.46$\pm$0.06(2) & $-0.26$ & 1.95 & -          & - & - & 0.43$\pm$0.07(2)        & $-0.29$ & 2.05  \\            
Nd II	& 60 &	1.42 & 1.26$\pm$0.17(11)     & $-0.16$ & 2.05& 0.59$\pm$0.12(8) & $-0.83$ & 1.62 & 0.90$\pm$0.20(5)    & $-0.52$ & 1.82 \\
Sm II	& 62 &	0.96 & 1.05$\pm$0.15(3)     & 0.09    & 2.30& 0.40$\pm$0.18(5)   & $-0.56$ & 1.89 &  0.99(1)     & 0.03    & 2.37 \\            
Eu II & 63 &  0.52 & 0.26$\pm$0.20(1,syn)         & $-0.26$ & 1.95 &  $-0.59$$\pm$0.20 (1,syn)     & $-1.11$      &  1.34 & $-1.36$$\pm$0.20 (1,syn)  & $-1.88$ & 0.46  \\
\hline
\end{tabular}
}

$^a$  \cite{asplund.2009ARA&A..47..481A}, The number inside the  parenthesis shows the number of lines used for the abundance determination. 
\label{table8}
\end{table*}
} 

{\footnotesize
\begin{table*}
\caption{\bf Estimates of [Fe/H], [ls/Fe], [hs/Fe], [hs/ls] and $^{12}$C/$^{13}$C}
\begin{tabular}{lccccc}
\hline                       
Star name    & [Fe/H]  & [ls/Fe] & [hs/Fe] & [hs/ls] & $^{12}$C/$^{13}$C\\ 
\hline
HE 0110$-$0406 & $-1.30$ & 1.03    & 1.36 & 0.33    & 45\\
HE 1425$-$2052 &$-$2.10  & 1.45    & 2.49 &  1.04   & 8.0 \\    
HE 1428$-$1950 & $-2.13$ & 1.14    & 1.07 & 0.07    & 23.0 \\
HE 1429$-$0551 & $-2.45$ & -       & 1.66 & -       & 20.0  \\  
HE 1447+0101   & $-2.25$ & -       & 2.03 & -       & 18.0 \\
HE 1523$-$1155 &$-2.44$  & 0.80    & 1.65 & 0.85    & 5.5  \\
HE 1528$-$0409 &$-2.35$  & -       & 1.69 &-        & 7.0  \\
\hline
\end{tabular}
\label{table9}
\end{table*}
}

{\footnotesize
\begin{table*}
\centering
\caption{\bf Comparison of the abundances of our programme stars with the literature values. }
\begin{tabular}{lccccccc}
\hline                       
Star name       & [Fe I/H]   & [Fe II/H]  & [Fe/H]     & [Sr/Fe]   & [Y/Fe]     & [Ba II/Fe] &  Ref \\
\hline
HE 1429$-$0551  & $-2.47$    & $-2.44$    &             &   -       &  -         &  1.88      & 1 \\ 
 & -          & -          & $-2.70$    &   -       &  0.89      &  1.52      & 2 \\ 
& $-2.47$    & $-2.48$    & $-2.47$    &   -       &  -         &  1.57      & 3 \\ 
HE 1447+0102    & $-2.29$    & $-2.21$    & $-2.25$    &   1.57    &  -         &  1.93      & 1\\
& $-2.47$    & $-2.45$    & $-2.46$    &   -       &  -         &  2.70      & 3\\
HE 1523$-$1155  & $-2.43$    & $-2.44$    & $-2.43$    &   0.93    &  0.68      &  1.83      & 1\\
& $-2.15$    & $-2.17$    & $-2.16$    &   -       &  -         &  1.72      & 3\\
HE 1528$-$0409  & $-2.37$    & $-2.34$    & $-2.35$    &   -       &  -         &  1.71      & 1\\
& $-2.61$    & $-2.59$    & $-2.60$    &   -       &  -         &  2.30      & 3\\
\hline                       
Star name       & [La II/Fe] & [Ce II/Fe] & [Pr II/Fe] & [NdII/Fe] & [Sm II/Fe] & [Eu II/Fe] &  Ref \\
\hline
HE 1429$-$0551  & 1.14       & 1.90       &   1.09     &  1.74     &  1.77      & 1.25       & 1 \\
 & 1.30       & 1.42       &   1.38     &  1.43     &  1.27      & 1.08       & 2 \\
\hline   
\end{tabular}

1. This work, 2. \cite{karinkuzhi.2021A&A...645A..61K}, 3. \cite{aoki.2007ApJ...655..492A}
\label{table10}
\end{table*}
} 

\section{\bf Abundance uncertainties}\label{sec:error_analysis}
The total uncertainties in the derived elemental abundances are contributed by random errors and systematic errors. While the uncertainties in the line parameters such as equivalent widths, line blending, and oscillator strength cause random errors, the uncertainties in the stellar atmospheric parameters causes systematic errors. We have followed the procedures as described  in \cite{shejeela.2021MNRAS.502.1008S} to calculate the uncertainties on the elemental abundances derived from the equivalent width measurement as well as the spectrum synthesis calculation. The total uncertainties in the estimated elemental abundances, log$\epsilon$ : 
\[\sigma^{2}_{log\epsilon} = \sigma^{2}_{ran}+ (\frac{\partial log\epsilon}{\partial T})^{2} \sigma_{T_{eff}}^{2}+ (\frac{\partial log\epsilon}{\partial logg})^{2}\sigma^{2}_{log g}+  (\frac{\partial log\epsilon}{\partial \zeta})^{2}\sigma^{2}_{\zeta}+ (\frac{\partial log\epsilon}{\partial [Fe/H]})^{2}\sigma^{2}_{[Fe/H])}~~~~~~~~~~(1)\]      

where $\sigma^{2}_{ran}$ = $\sigma_{s}/\sqrt{N}$ . $\sigma_{s}$ represents the standard deviation in the elemental abundances derived using N number of lines due to that element. The $\sigma$’s indicate the typical uncertainties in the adopted stellar atmospheric parameters, and are T$_{eff}$ $\sim$ $\pm$ 100 K, log g $\sim$ $\pm$ 0.2 dex, $\zeta$ $\sim$ $\pm$ 0.2 km s$^{-1}$ , and [Fe/H] $\sim$ $\pm$ 0.1 dex. The uncertainty in [X/Fe] is determined using the relation :
\[\sigma^{2}_{[X/Fe]} = \sigma^{2}_{X} + \sigma^{2}_{[Fe/H]}~~~~~~~~~~(2) \]
As an example, the differential elemental abundances derived using the equation (2) for the object HE 1523-1155 are given in Table \ref{table11}. The error for the abundances derived from the spectrum synthesis calculation is taken to be 0.2 dex (indicated by “syn” in Tables \ref{table6}, \ref{table7}, \ref{table8}). This is the minimum change in the abundance value producing well distinguished synthetic spectra with respect to the best fits. For abundances derived using the equivalent width measurements, the errors listed represent the standard deviation (Tables \ref{table6}, \ref{table7}, \ref{table8}).

{\footnotesize
\begin{table*}
\centering
\caption{\bf Differential elemental abundances (log$\epsilon$) derived for the object HE 1523$-$1155. }
\begin{tabular}{lccccccc}
\hline                       
element	& $\Delta$ T$_{eff}$ &	$\Delta$ log g & $\Delta \zeta$ &	$\delta$ [Fe/H] &	($\sum$ $\sigma^{2}_{i}$)$^{1/2}$ &	$\sigma$[X/Fe]\\
 &($\pm$100K)&	($\pm$0.2dex)&	($\pm$0.2kms$^{-1}$) &	($\pm$0.1 dex) &  &\\	
\hline
C	&	$\pm$0.21	&	$\mp$0.03	&	0.00	&	0.00	&	0.21	& 0.31		\\
N	&	$\pm$0.31	&	$\mp$0.07	&	$\pm$0.02	&	$\pm$0.04	&	0.32	& 0.39		\\
O	&	$\pm$0.05	&	$\pm$0.06	&	0.00	&	$\pm$0.02	&	0.08	& 0.24		\\
Na I 	&	$\pm$0.15	&	$\mp$0.08	&	$\mp$0.08	&	$\mp$0.02	&	0.19	&	0.22	\\
Mg I	&	$\pm$0.07	&	$\mp$0.05	&	$\mp$0.09	&	$\mp$0.01	& 0.12	&	0.17	\\
Ca I	&	$\pm$0.08	&	$\mp$0.02	&	$\mp$0.05	&	0.00	&	0.10	&	0.15	\\
Sc II	&	$\pm$0.06	&	$\pm$0.08	&	0.00	&	$\pm$0.02	&	0.10	&	0.25	\\
Ti I	&	$\pm$0.14	&	$\mp$0.03	&	$\mp$0.04	&	$\mp$0.01	&	0.15	&	0.19	\\
Ti II	&	$\pm$0.05	&	$\pm$0.05	&	$\mp$0.08	&	0.00	&	0.11	&	0.16	\\
V I	&	$\pm$0.12	&	$\mp$0.03	&	$\mp$0.05	&	$\mp$0.01	&	0.13	&	0.26	\\
Cr I	&	$\pm$0.13	&	$\mp$0.02	&	$\mp$0.08	&	$\mp$0.01	&	0.15	&	0.22	\\
Cr II	&	$\mp$0.01	&	$\pm$0.07	&	$\mp$0.01	&	$\pm$0.01	&	0.07	&	0.13	\\
Fe I	& $\pm$0.11		& $\pm$0.01		&	$\pm$0.04	&	$\mp$0.04	& 0.12		&		\\
Fe II	&	$\pm$0.04	&	$\pm$0.09	&	$\mp$0.03	& $\pm$0.01		& 0.10	&		\\
Ni I	&	$\pm$0.11	&	$\mp$0.01	&	$\mp$0.03	&	$\mp$0.01	& 0.11	&	0.17	\\
Zn I	&	$\pm$0.04	&	$\pm$0.03	&	$\mp$0.02	&	$\pm$0.01	& 0.05	&	0.12	\\
Sr I	&	$\pm$0.30	&	$\pm$0.08	&	$\pm$0.08	&	$\pm$0.05	& 0.32	&	0.40	\\
Y II	&	$\pm$0.06	&	$\pm$0.06	&	$\mp$0.12	&	$\pm$0.01	& 0.15	&	0.18	\\
Ba II	&	$\pm$0.08	&	$\mp$0.04	&	$\mp$0.25	&	$\mp$0.09	& 0.28	&	0.36	\\
La II	&	$\pm$0.07	& $\pm$0.05	&	$\mp$0.06	&	0.00	&	0.10	& 0.25	\\
Ce II	&	$\pm$0.07	&	$\pm$0.06	&	$\mp$0.06	&	$\pm$0.01	&	0.11	&	0.17	\\
Nd II	&	$\pm$0.08	&	$\pm$0.06	&	$\mp$0.05	&	$\pm$0.01	&	0.11	&	0.16	\\
Sm II	&	$\pm$0.07	&	$\pm$0.06	&	$\mp$0.04	&	$\pm$0.01	&	0.10	&	0.17	\\
Eu II	&	$\pm$0.04	&	$\pm$0.06	&	$\pm$0.02	&	$\pm$0.02	&	0.08	&	0.24	\\
\hline   
\end{tabular}
\label{table11}
\end{table*}
}

\section{Kinematic analysis}\label{sec:kinematic_analysis}

Kinematic analysis of the programme stars are performed following a detailed procedure as explained in \cite{purandardas.2021arXiv210307075P}. We have used right-handed coordinate system for the determination of space velocities of the programme stars. That is, the three components U, V and W of the space velocities, U is positive in the direction of Galactic center, V is positive in the direction Galactic rotation and W is positive in the direction of the North Galactic Pole  \citep{johnson&soderblom.1987AJ.....93..864J}. The values of proper motions and parallax required for the calculations of space velocities are taken from SIMBAD and \textit{Gaia} \citep{gaia.2016A&A...595A...1G,gaia.2018A&A...616A...1G} respectively. We have used our radial velocity estimates for the calculations.

\par We have determined the probability for a star's membership into the thin disk, the thick disk or the halo population. Our estimates show that the objects HE 0110$-$0406 and HE 1447+0102 belong to thick disk population with a probability of $\sim$ 94\% and 84\% respectively. The remaining objects belong to the Halo population with a probability $>$ 81\%. The spatial velocities and the probability estimates for the programme stars are presented in Table \ref{table12}. 

{\footnotesize
\begin{table*}
\caption{\bf Spatial velocity and probability estimates for the programme stars}
\resizebox{0.9\textwidth}{!}{\begin{tabular}{lcccccccc}
\hline                       
Star name           & U$_{LSR}$         & V$_{LSR}$           & W$_{LSR}$ & V$_{spa}$  & p$_{thin}$ & p$_{thick}$ & p$_{halo}$ & Population \\
                    & (kms$^{-1}$)        & (kms$^{-1}$)          & (kms$^{-1}$) &  (kms$^{-1}$) &           &           & &  \\  
    \hline
HE 0110$-$0406 & 66.27$\pm$4.91      & $-13.36$$\pm$1.88    & 29.28$\pm$3.96       & 73.67$\pm$5.64    & 0.94 & 0.06 & 0.00 & Thick \\  
HE 1425$-$2052   & 289.78$\pm$100.54   & $-156.13$$\pm$63.38  & $-230.36$$\pm$161.09 & 401.77$\pm$45.33  & 0.00 & 0.00 & 1.00 & Halo \\
HE 1428$-$1950   & 135.01$\pm$17.25    & $-131.24$$\pm$25.52  & $-128.65$$\pm$35.87  & 228.04$\pm$24.33  & 0.00 & 0.18 & 0.81 & Halo \\
HE 1429$-$0551   & 7.65$\pm$5.19       & $-501.65$$\pm$97.37  & $-186.93$$\pm$29.35  & 535.40$\pm$101.39 & 0.00 & 0.00 & 1.00 & Halo \\
HE 1447+0102   & $-26.39$$\pm$0.74   & $-159.84$$\pm$16.69  & $-53.82$$\pm$1.36    & 170.71$\pm$16.17  & 0.00 & 0.84 & 0.16 & Thick \\
HE 1523$-$1155   & $-128.62$$\pm$14.75 & $-307.01$$\pm$44.54  & 66.61$\pm$11.44      & 339.46$\pm$43.54  & 0.00 & 0.00 & 0.99 & Halo \\
HE 1528$-$0409   & 117.61$\pm$73.27    & $-290.93$$\pm$100.23 & $-353.23$$\pm$86.61  & 472.48$\pm$105.49 & 0.00 & 0.00 & 1.00 & Halo\\
\hline
\end{tabular}}
\label{table12}
\end{table*}
}

\section{Discussion}\label{sec:discussions}
We have carried out a detailed abundance analysis for a sample of  seven objects, one object is found to be  metal-poor and the remaining six are found to be very metal-poor stars. Our analysis shows that the objects HE 1447+0102 and HE 1523$-$1155 satisfy the criteria for  CEMP-r/s stars ([C/Fe] $>$ 0.7 \citep{aoki.2007ApJ...655..492A}, [Ba/Fe] $\geq$ 1.0, [Eu/Fe] $\geq$ 1.0, 0.0 $\leq$ [Ba/Eu] $\leq$ 1.0 and/or 0.0 $\leq$ [La/Eu] $\leq$ 0.7 \citep{partha.2021A&A...649A..49G}) and all the remaining objects satisfy the criteria for CEMP-s stars ([C/Fe] $>$ 0.7 \citep{aoki.2007ApJ...655..492A}, [Ba/Fe] $\geq$ 1.0, [Ba/Eu] $>$ 0.0 \citep{partha.2021A&A...649A..49G}). We could not classify HE 1428-1950 based on criteria involving Ba abundance  as we could not determine the abundance of Ba in this object. 

\par All the programme stars exhibit enhancement of carbon and nitrogen except HE 1425$-$2052 for which nitrogen is only slightly enhanced. Oxygen is also found to be enhanced in the stars except HE 1523$-$1155 and HE 1528$-0409$. The $\alpha$-elements are found to be slightly enhanced in all the programme stars except HE 0110$-$0406 for which the abundance of $\alpha$-elements are found to be near solar. While HE 1425$-$2052 shows near solar abundance of Fe-peak elements, HE 1428$-$1950 exhibits enhancement of Fe-peak elements except for Ti and Sc. The remaining objects show slight enhancement in Fe-peak elements. 

\par
\begin{figure}
\centering
\includegraphics[width=12cm,height=11cm]{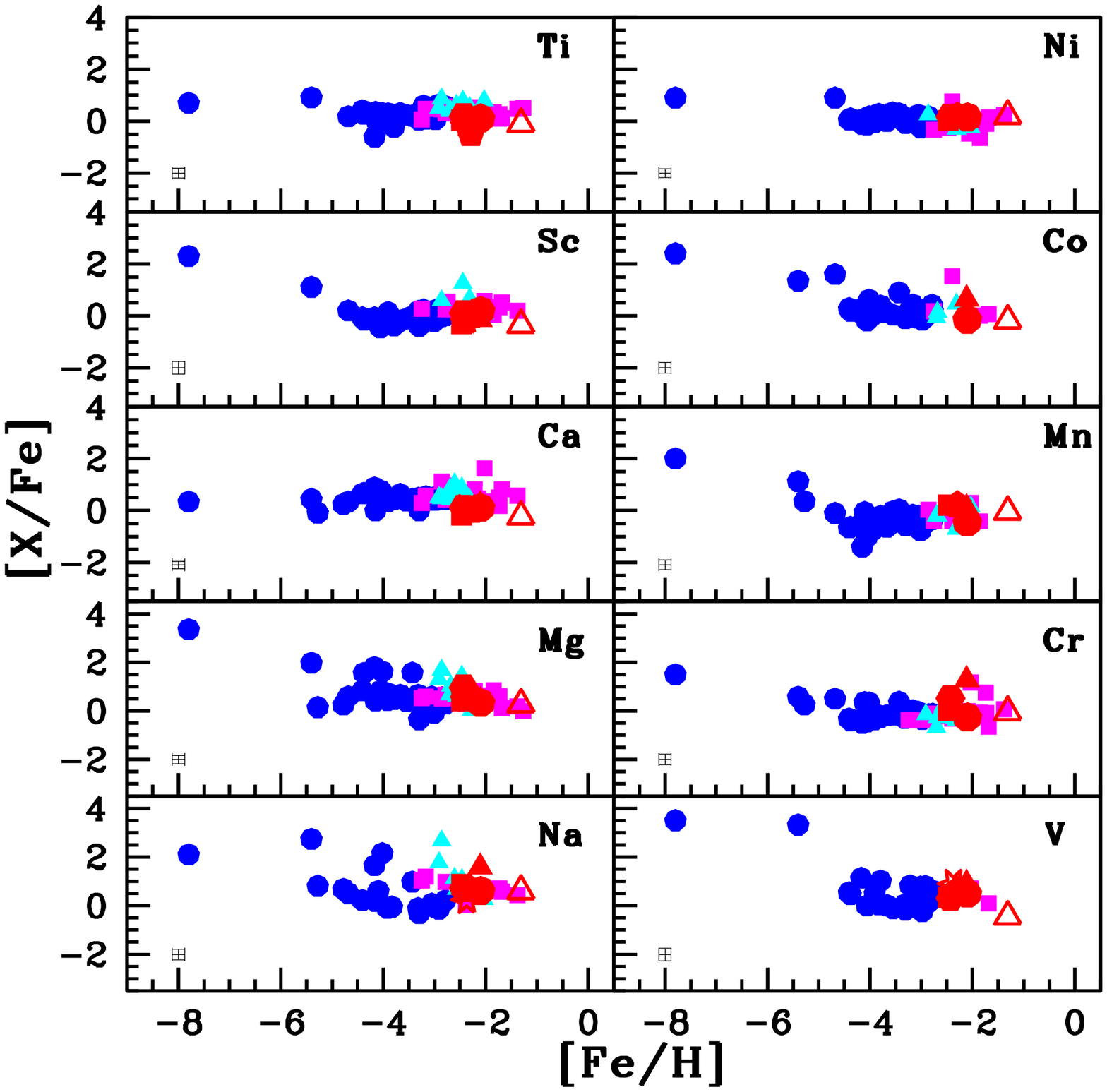}
\caption{Abundance ratios of light elements with respect to metallicity estimated in our programme stars are represented using red color symbols: HE 0110$-$0406 is represented by open triangle. Filled circle represents HE 1425$-$2052. Filled triangle indicates HE 1428$-$1950. HE 1429$-$0551 is represented by filled square. Filled pentagon represents HE 1447+0102. Filled hexagon and asterisk symbol indicate HE 1523$-$1155 and HE 1528$-$0409 respectively. CEMP-s and CEMP-r/s stars are represented by filled squares (magenta) and filled triangles respectively. Filled circles(blue) represent CEMP-no stars. The abundance values for CEMP-no stars used for the comparison are taken from \cite{christlieb.2004A&A...428.1027C}, \cite{plez.cohen.2005A&A...434.1117P}, \cite{yong.2013ApJ...762...26Y}, \cite{hansen.2014ApJ...787..162H}, \cite{bonifacio.2015A&A...579A..28B}, \cite{bessel.2015ApJ...806L..16B} and \cite{frebel.2018ARNPS..68..237F}. The abundance
values for CEMP-s and CEMP-r/s stars are taken from \cite{lucatello.2003AJ....125..875L}, \cite{barklem.2005A&A...439..129B}, \cite{cohen.2006AJ....132..137C}, \cite{goswami.2006MNRAS.372..343G},
\cite{aoki.2007ApJ...655..492A}, \cite{karinkuzhi.2015MNRAS.446.2348K}, \cite{purandardas.2019aMNRAS.486.3266P}, \cite{purandardas.2019bBSRSL..88..207P} and \cite{partha.2021A&A...649A..49G}. Representative error bar is shown at the left bottom corner of the figure.}
\label{fig7}
\end{figure}

\begin{figure}
\centering
\includegraphics[width=12cm,height=11cm]{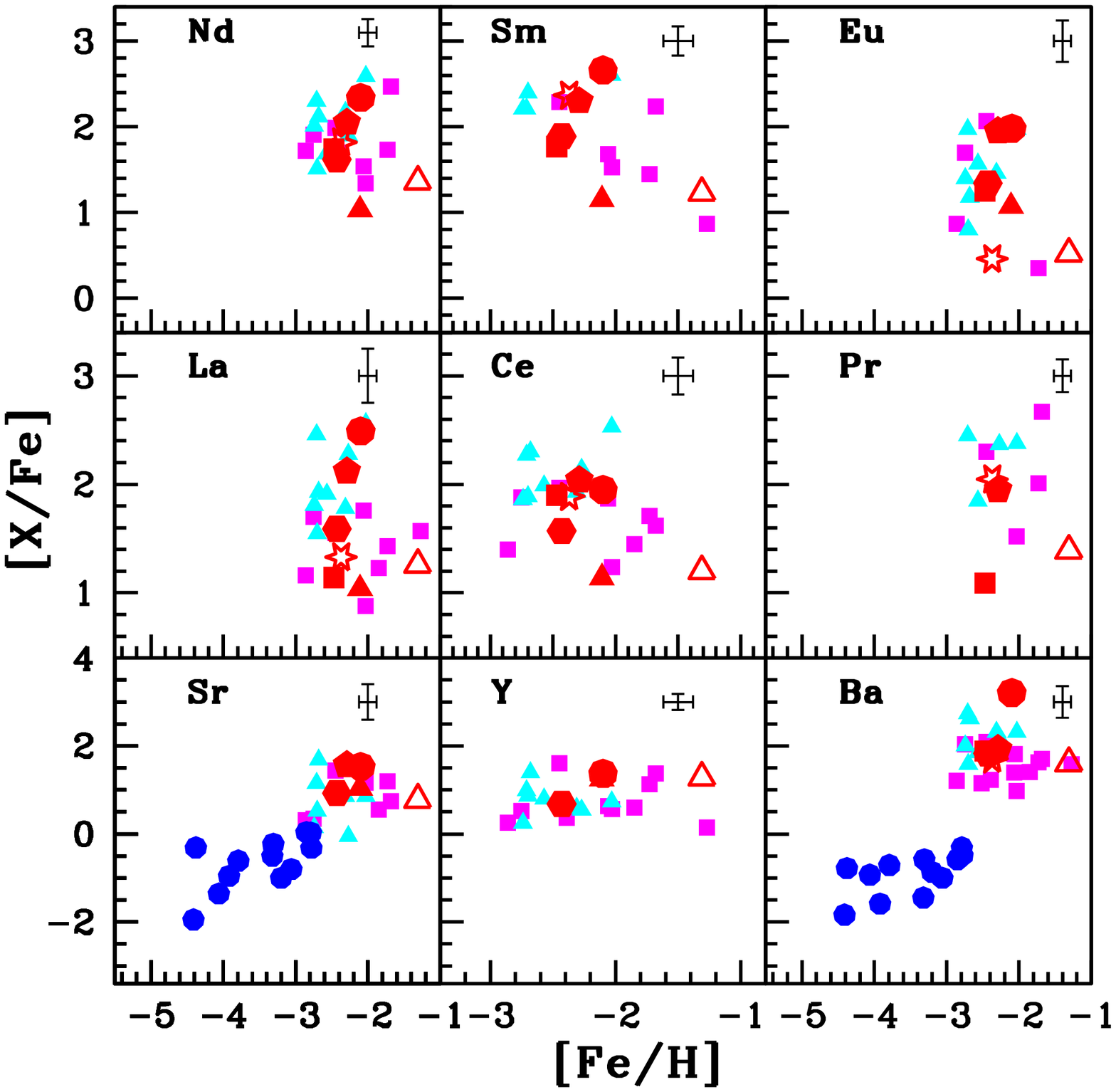}
\caption{Same as Figure \ref{fig7}, but for heavy elements. Representative error bar is shown at the right top corner of the figure.}
\label{fig8}
\end{figure}

 A comparison of the abundances of various elements estimated in our programme stars with their counterparts observed in CEMP-s, CEMP-r/s and CEMP-no stars taken from various sources is shown in the Figures \ref{fig7} and \ref{fig8}. Although CEMP-s and r/s stars fall at slightly higher metallicity range than CEMP-no stars, all of them exhibit similar $\alpha$ and Fe peak elemental abundance patterns except for Mn and V. A substantial number of CEMP-no stars show [Mn/Fe] and [V/Fe] $<$ 0 unlike CEMP-s and CEMP-r/s stars. Majority of the CEMP-no stars exhibit Na abundance in the range $-$0.50 $<$ [Na/Fe] $<$ 0.50, and  CEMP-s and r/s stars show [Na/Fe] $>$ 0. Unlike CEMP-s and CEMP-r/s stars, the heavy elements such as Sr and Ba are found to be underabundant in CEMP-no stars.

\par The location of a star in the absolute carbon  abundance A(C) vs. metallicity [Fe/H] diagram,  and various elemental abundance ratios can give important clues about the possible origin of the observed enhancement of carbon as well as neutron-capture elements. A discussion along this line is presented  in detail in the following subsections, in the context of the programme stars.

\subsection{Locations of the programme stars in the A(C) vs. [Fe/H] diagram:}

Many authors have shown that the CEMP stars exhibit bimodal distribution in the A(C) vs. [Fe/H] diagram \citep{spite.2013A&A...552A.107S,bonifacio.2015A&A...579A..28B,hansen.2015ApJ...807..173H,yoon.2016ApJ...833...20Y}. While \cite{spite.2013A&A...552A.107S} claimed a higher carbon band around A(C) $\sim$ 8.25 and a lower carbon band around A(C) $\sim$ 6.5, \cite{yoon.2016ApJ...833...20Y} suggested a higher carbon band which peaks around A(C) $\sim$ 7.96 and a lower band peaking at A(C) $\sim$ 6.28. \cite{yoon.2016ApJ...833...20Y} further classified the objects in the A(C) vs. [Fe/H] diagram into three groups. In this diagram, the objects that show very weak dependence of A(C) on [Fe/H] are called as Group I objects. This group is distributed around the higher carbon band. They are mostly composed of CEMP-s and CEMP-r/s stars and a large fraction of them are confirmed binaries. Group II and Group III objects are extremely metal-poor objects, mostly composed of CEMP-no stars, that are clustered around the lower carbon band.  While for Group III objects, A(C) shows no clear dependence on [Fe/H], the A(C) values for Group II objects exhibit very clear dependence on [Fe/H]. A large fraction of Group II and Group III objects are found to be single and hence the observed abundance anomalies are believed to be intrinsic in origin.

\par
\begin{figure}
\centering
\includegraphics[width=12cm,height=11cm]{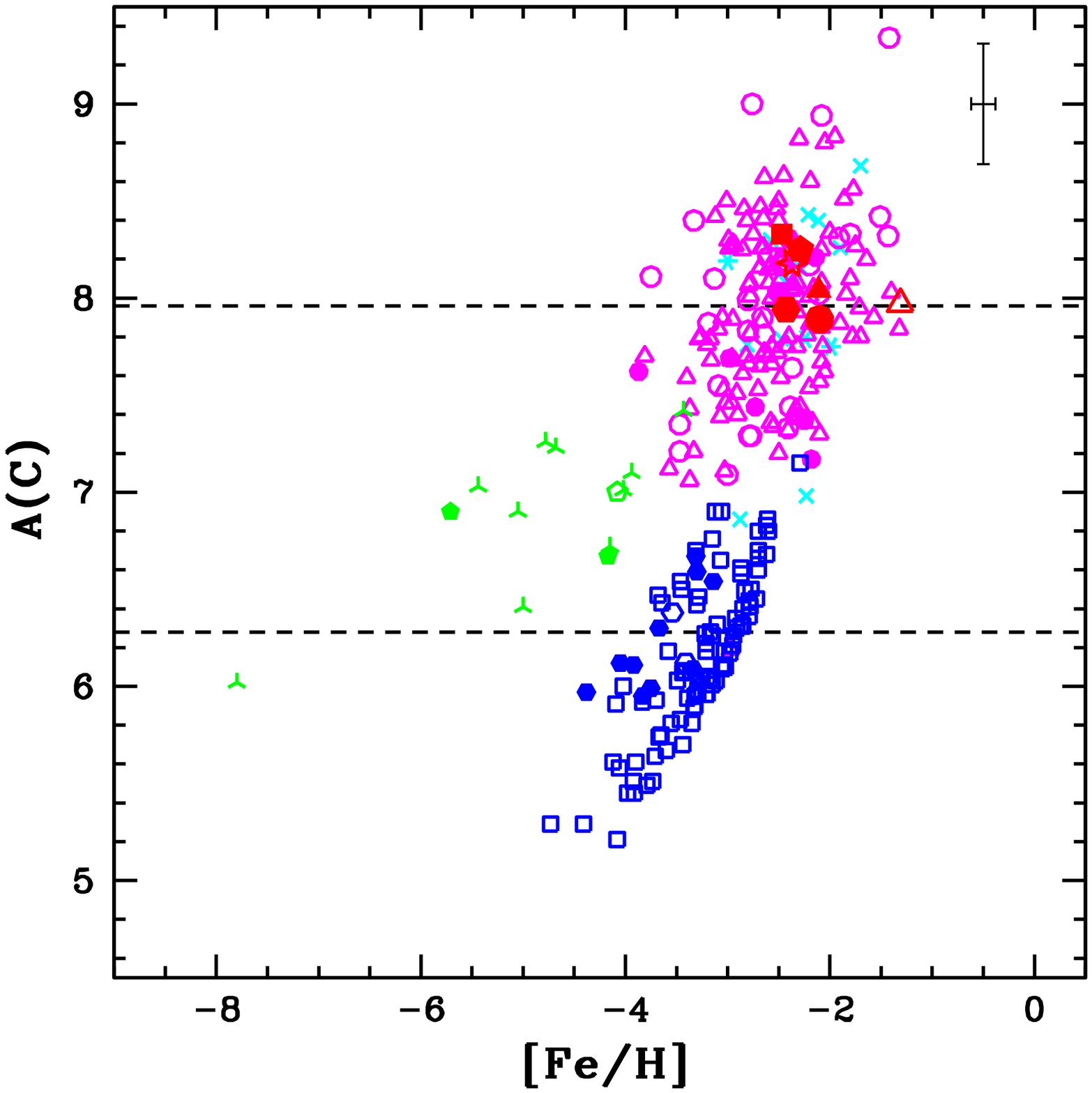}
\caption{Corrected A(C) vs. [Fe/H] diagram for the compilation of CEMP stars taken from \cite{yoon.2016ApJ...833...20Y}. Cyan symbols indicate CEMP-r/s stars: binary stars are represented by eight sided star and stars with no information about the binary status are indicated by cross symbols. CEMP-s stars are represented using magenta symbols: Open and filled circles indicates binary and single stars respectively. Stars with no information about the binarity are represented using open triangle. Blue symbols represent Group II CEMP-no stars: Open and filled hexagons represent binary and single stars respectively. Stars with no information about the binarity are presented using open square. Group III CEMP-no stars are represented using green symbols: binary and single stars are represented by open and filled pentagons respectively. Skeleton triangle represents stars with no information about binary status. The symbols used for our programme stars are same as in Figure \ref{fig7}. Representative error bar is shown at the right top corner of the figure.}
\label{fig9}
\end{figure}

 The locations of our programme stars in the A(C) vs. [Fe/H] diagram is shown in Figure \ref{fig9}. In order to locate the programme stars in the A(C) vs. [Fe/H]  diagram of \cite{yoon.2016ApJ...833...20Y}, corrections have to be applied to the estimated abundance of carbon. The necessary corrections are applied
using the public online tool by \cite{placco.2014bApJ...797...21P} 
\footnote {\url{http://vplacco.pythonanywhere.com/}}. 
Table \ref{table13} presents the carbon correction and corrected carbon abundances obtained for our programme stars.
All of our programme stars are found to be Group I objects in the A(C) vs. [Fe/H] diagram. We therefore expect that the observed enhancement of carbon and heavy elements in these stars may be attributed to a binary companion. But none of the programme stars are confirmed binaries except HE 1523$-$1155.

\begin{table}
\caption{\bf Carbon correction and the corrected carbon abundances obtained for our programme stars}
\begin{tabular}{lcc}
\hline
 star name   & Carbon correction & log$\epsilon$(C)$_{c}$\\
\hline
HE 0110$-$0406 &   0.12 &  7.97  \\
HE 1425$-$2052 &  0.13 & 7.89 \\
HE 1428$-$1950 &  0.12 & 8.04 \\
HE 1429$-$0551 &  0.08 & 8.33 \\
HE 1447+0102   &  0.06 & 8.25 \\
HE 1523$-$1155 &  0.11 & 7.94 \\
HE 1528$-$0409 &  0.05 & 8.18 \\
\hline
\end{tabular}
\label{table13}

Note : log$\epsilon$(C)$_{c}$ represents the carbon abundance after applying the correction.
\end{table}

\subsection{Possible progenitors of the programme stars}

The enhancement of heavy elements in CEMP-s and CEMP-r/s stars are extrinsic in origin as the evolutionary status of these objects do not support such enhancements. A brief discussion on CEMP-s and CEMP-r/s stars is presented in section \ref{sec:intro}. As discussed earlier, i-process can well explain the observed over abundances of s- and r-process elements in CEMP-r/s stars. It is found that the neutron densities required for i-process can be achieved in AGB stars during proton ingestion episodes \citep{cowan&rose1977ApJ...212..149C,campbell.2008A&A...490..769C,cristallo.2009PASA...26..139C,campbell.2010A&A...522L...6C,stancliffe.2011ApJ...742..121S,banerjee.2018MNRAS.480.4963B,clarkson.2018MNRAS.474L..37C}. According to this scenario, a CEMP-r/s star accrets material synthesized by its binary companion during its AGB phase of evolution by i-process nucleosynthesis that can enrich its surface compositions by s- and r- process elements. 

\par
\begin{figure}
\centering
\includegraphics[width=12cm,height=11cm]{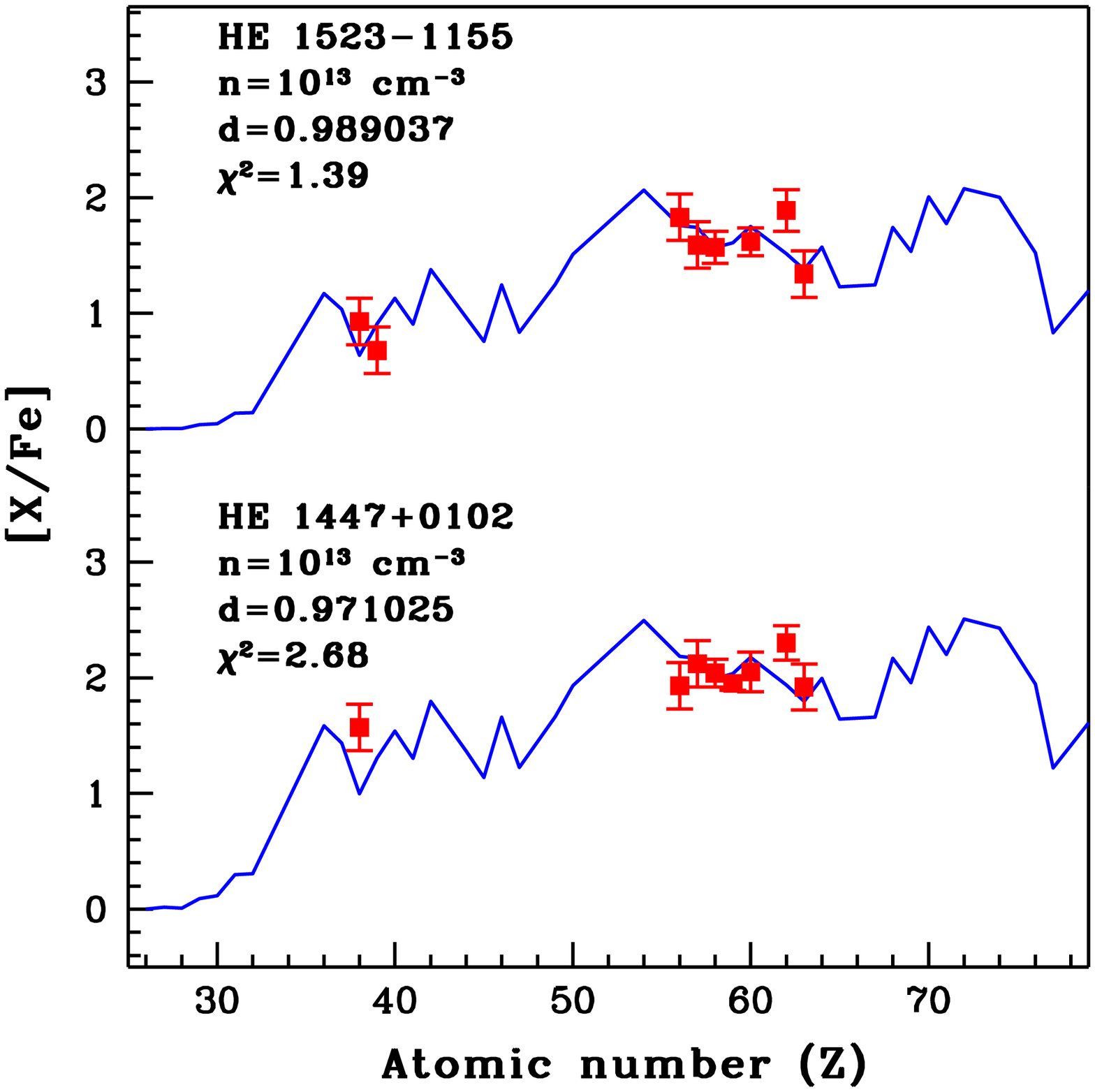}
\caption{Best-fitting i-process model (solid blue curve) for HE 1447+0102 and HE 1523$-$1155. The points with error bars indicate the observed abundances.}
\label{fig10}
\end{figure}

 Abundance patterns in 20 CEMP-r/s stars could be reproduced by \cite{hampel.2016ApJ...831..171H} based on their i-process model which supports the i-process as the probable mechanism for the formation of CEMP-r/s stars. \cite{hampel.2016ApJ...831..171H} have calculated the yields of heavy elements from i-process nucleosynthesis occurring at different neutron densities based on single-zone nuclear network calculations. We have compared the abundance patterns estimated in the CEMP-r/s stars, HE 1447+0102 and  HE 1523$-$1155 with the model yields of \cite{hampel.2016ApJ...831..171H} as shown in Figure \ref{fig10}. The abundance of neutron-capture elements in HE 1447+0102 and HE 1523$-$1155 match well with the expected yields from the i-process nucleosynthesis at a neutron density n $\sim$ 10$^{13}$ with a dilution factor $\sim$ 0.98. 

\par
\begin{figure}
\centering
\includegraphics[width=8cm,height=8cm]{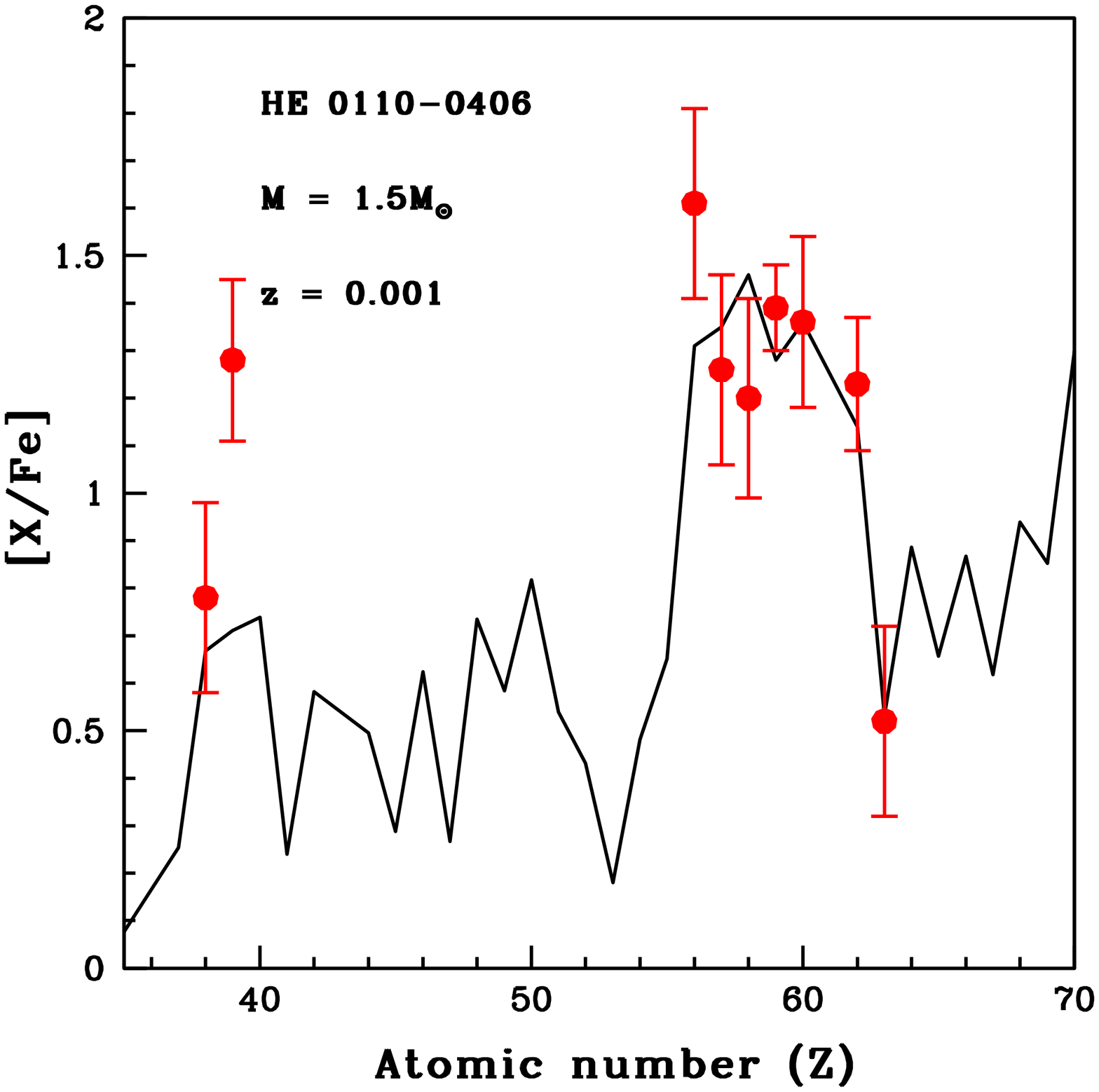}
\includegraphics[width=8cm,height=8cm]{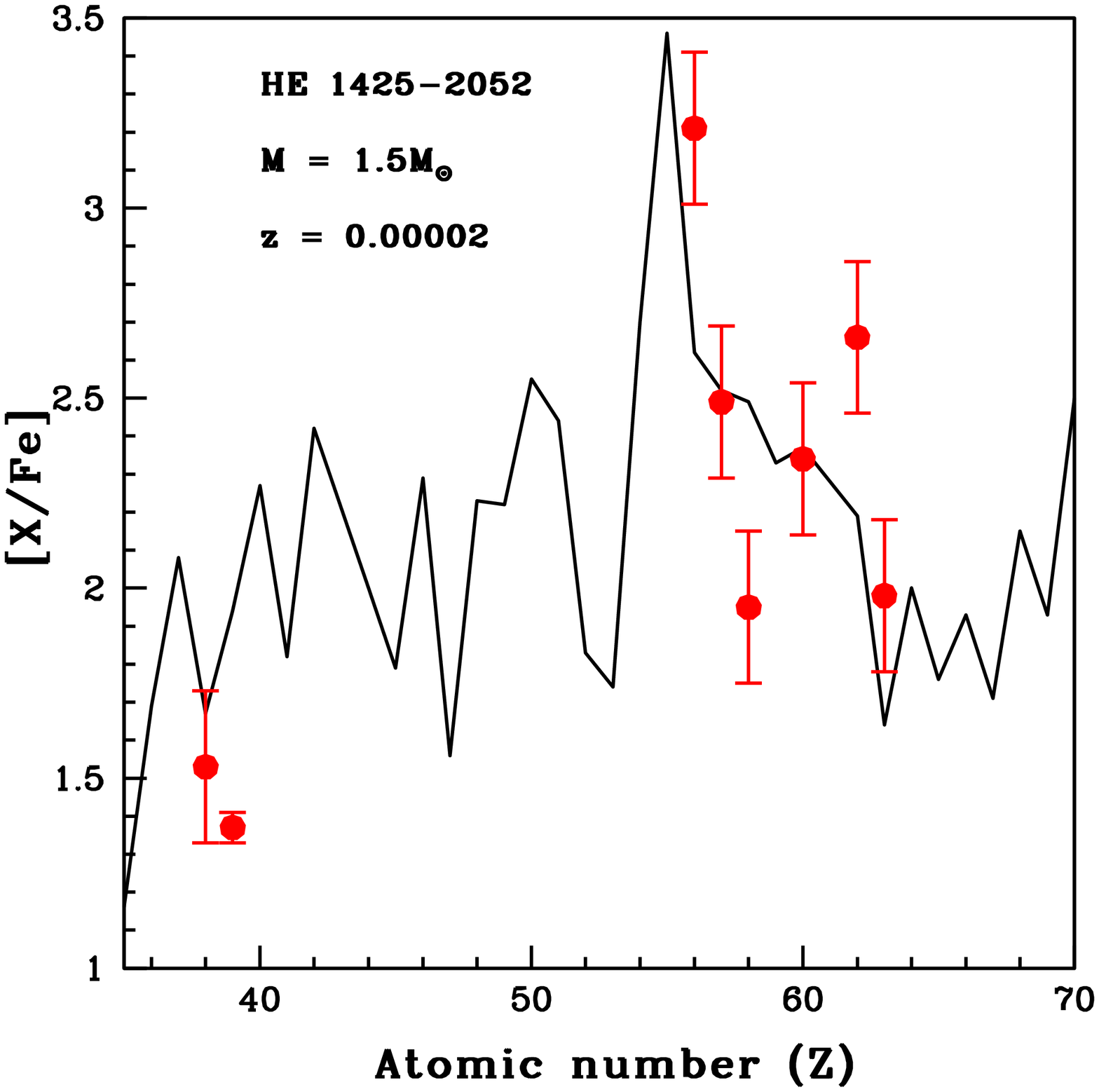}
\caption{Best-fitting FRUITY model (solid black curve) for HE 0110$-$0406, and HE 1425$-$2052. The points with error bars indicate the observed abundances.}
\label{fig11}
\end{figure}

\begin{figure}
\centering
\includegraphics[width=8cm,height=8cm]{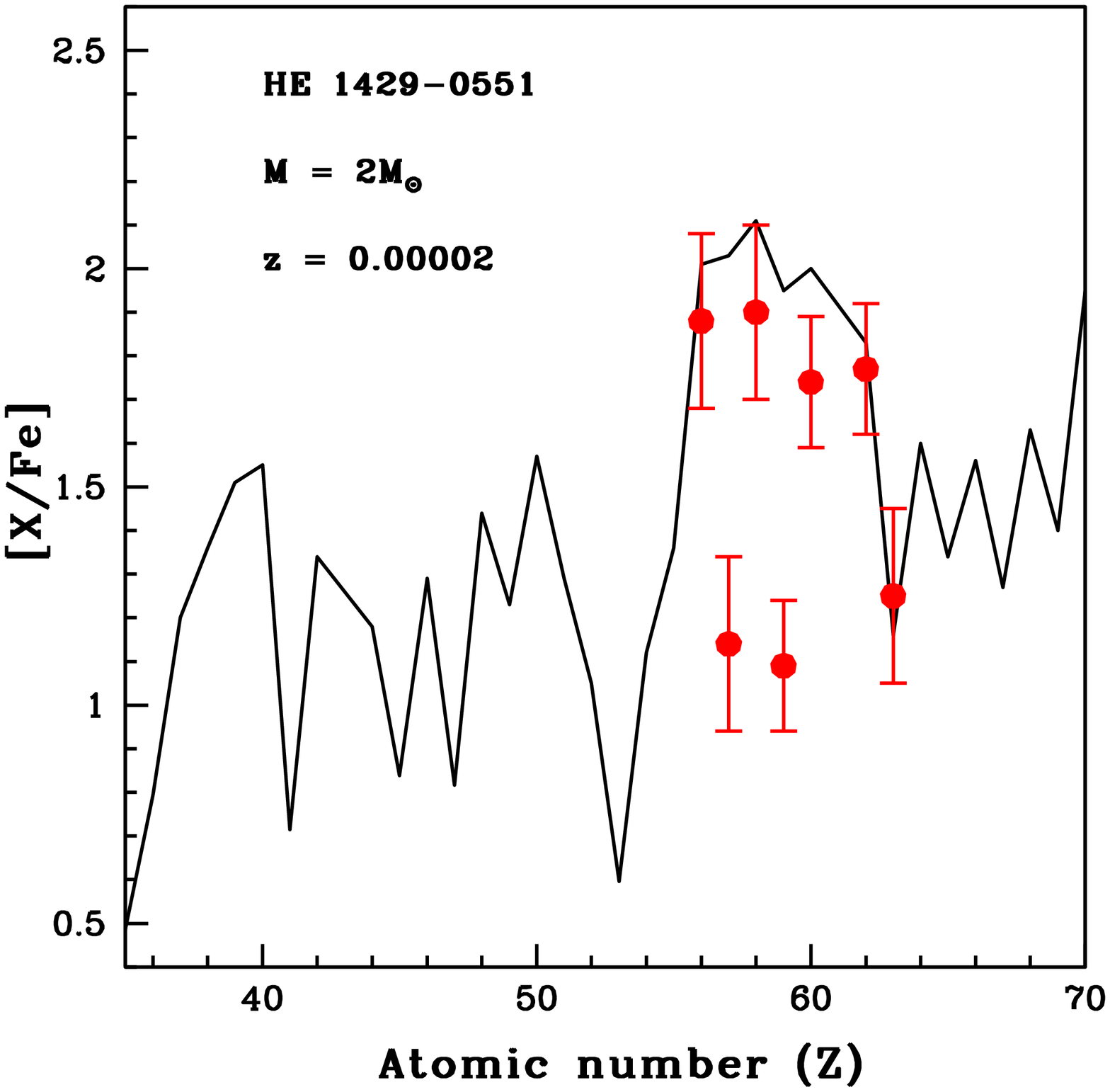}
\includegraphics[width=8cm,height=8cm]{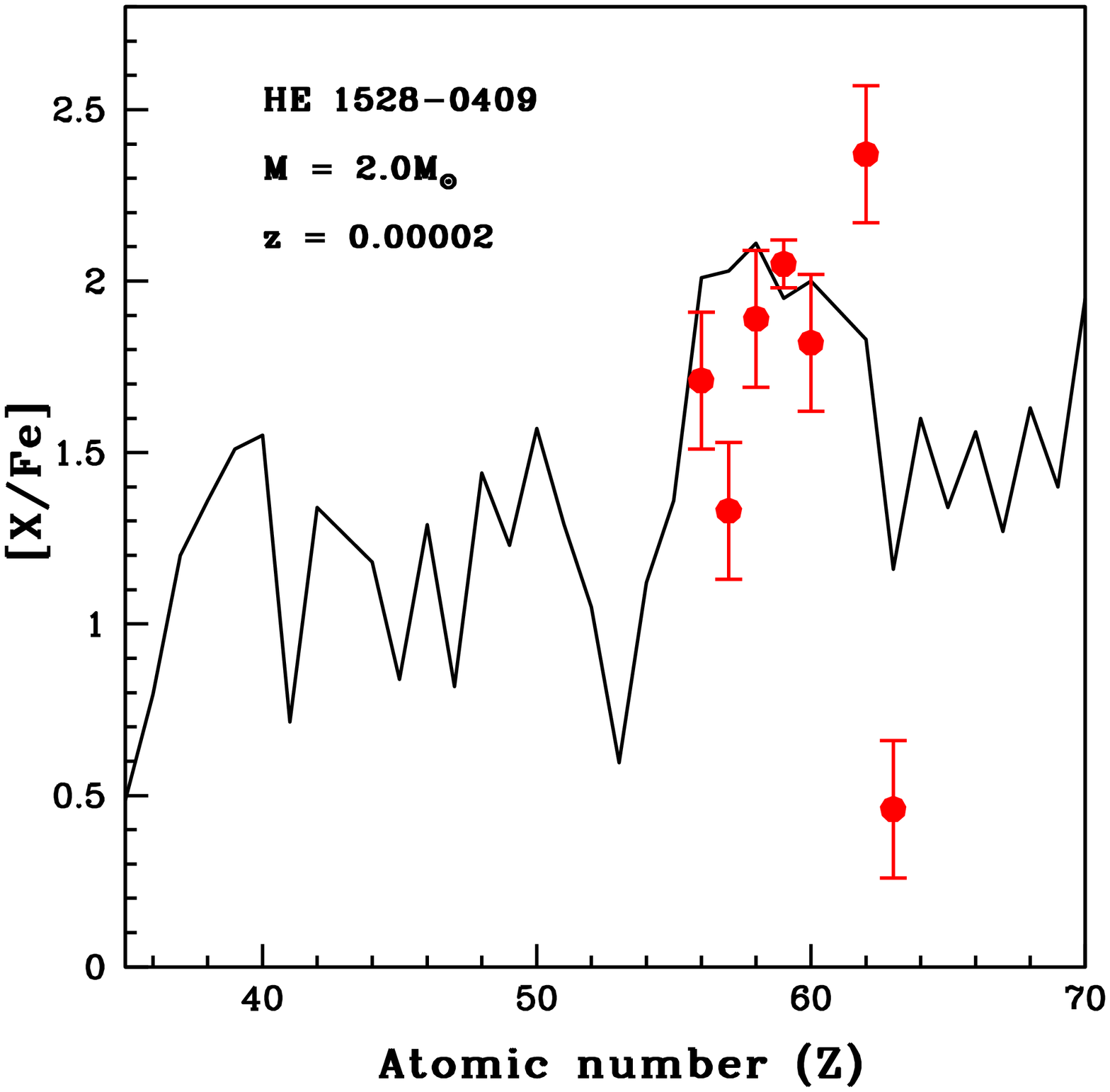}
\caption{Best-fitting FRUITY model (solid black curve) for HE 1429$-$0551, and HE 1528$-$0409. The points with error bars indicate the observed abundances.}
\label{fig12}
\end{figure}

 We have compared the abundance patterns of  CEMP-s stars in  our sample with the model predictions of  AGB stars provided by FRANEC Repository of Updated Isotopic Tables \& Yields (FRUITY) models \citep{cristallo.2009ApJ...696..797C,cristal.2011ApJS..197...17C,cristallo.2015ApJS..219...40C};  the data are publicly available at \url{http://fruity.oa-teramo.inaf.it/} following the  detailed procedure presented in \cite{shejeela.2021MNRAS.502.1008S} . The best fits obtained for the observed abundances are shown in Figures \ref{fig11} and \ref{fig12}. For the object HE 0110-0406, the best fit is obtained for the model with metallicity, z = 0.001, and mass, M = 1.5M$_{\odot}$. For HE 1429-0551, and HE 1528-0409, the model with z = 0.00002 and M = 2M$_{\odot}$ gives the best fit. For HE 1425-2052, the best fit is obtained for the model with z = 0.00002, and mass, M = 1.5M$_{\odot}$. The observed abundance patterns in HE 1428-1950 could not be matched with any of the available yield values from  the FRUITY models.

\par The abundance of oxygen and the [Sr/Ba] ratio can be used as indicators of the possible formation sites of a star \citep{choplin.2017A&A...607L...3C}. \cite{choplin.2017A&A...607L...3C} analysed the abundance patterns of four single CEMP-s stars based on the models of rotating and non-rotating massive stars with masses 10-150 M$_{\odot}$. They find that the models of fast rotating massive stars could reproduce the observed abundance patterns in three of the CEMP-s stars in their sample. Their studies show that CEMP-s stars with fast rotating massive star as the progenitor show oxygen abundance in the range between 1.5 to 2. \cite{karakas.2010MNRAS.403.1413K} predict $-0.2$ $<$ [O/Fe] $<$ 1.2 for CEMP-s stars with AGB progenitors. Studies by \cite{Frischknecht.2012A&A...538L...2F} show that [Sr/Ba] $>$ 0 for massive rotating stars. \cite{choplin.2017A&A...607L...3C} also support this idea and they claim that [Sr/Ba] ratio is higher in fast rotating massive stars than in AGB stars. 

\par
\begin{figure}
\centering
\includegraphics[width=12cm,height=11cm]{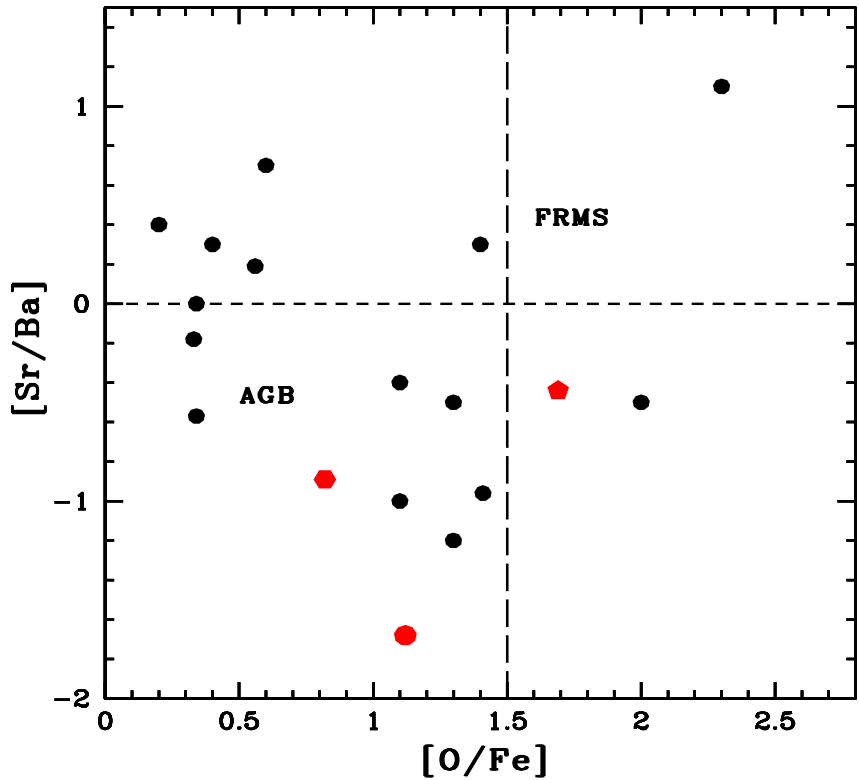}
\caption{Locations of the programme stars HE 1425$-$2052, HE 1447+0102 and HE 1523$-$1155 in [Sr/Ba] vs. [O/Fe] diagram. Symbols used for the programme stars are same as in Figure \ref{fig7}. Filled circles (black) represent CEMP stars from \cite{hansen.2019A&A...623A.128H}}
\label{fig13}
\end{figure}

 We could determine [Sr/Ba] ratio in HE 0110$-$0406, HE 1425$-$2052, HE 1447+0102 and HE 1523$-$1155. For all of them the [Sr/Ba] ratio is found to be $<$ 0. This indicates an AGB star as  the source of the peculiar abundance patterns in these stars. While the [Sr/Ba] ratio in HE 1447+0102 indicates an AGB star as the progenitor, the [O/Fe] measured in this object is the characteristic of a fast rotating massive star progenitor. This shows that this object may be polluted by multiple events. The [O/Fe] estimated in the remaining stars represent an AGB progenitor except HE 1429$-$0551 and HE 1447+0102. We could measure both [Sr/Ba] ratio and [O/Fe] only in three of the programme stars, HE 1425$-$2052, HE 1447+0102 and HE 1523$-$1155 and their locations in the [Sr/Ba] vs. [O/Fe] diagram are shown in Figure \ref{fig13}.

\par Estimation of [ls/hs] ratio is another way to find the possible progenitor of a star. While the models of fast rotating massive stars predict [ls/hs] $\geq$ 0 \citep{choplin.2017A&A...607L...3C,chiappini.2013AN....334..595C,cescutti.2013A&A...553A..51C}, the AGB models predict [ls/hs] $<$ 0 \citep{abate.2015A&A...581A..22A}. We could measure the [ls/hs] ratio in HE 0110$-$0406, HE 1425$-$2052, HE 1428$-$1950 and HE 1523$-$1155. For all of them, [ls/hs] ratio is found to be $<$ 0. This indicates an AGB progenitor for these objects.
   
\par
\begin{figure}
\centering
\includegraphics[width=12cm,height=11cm]{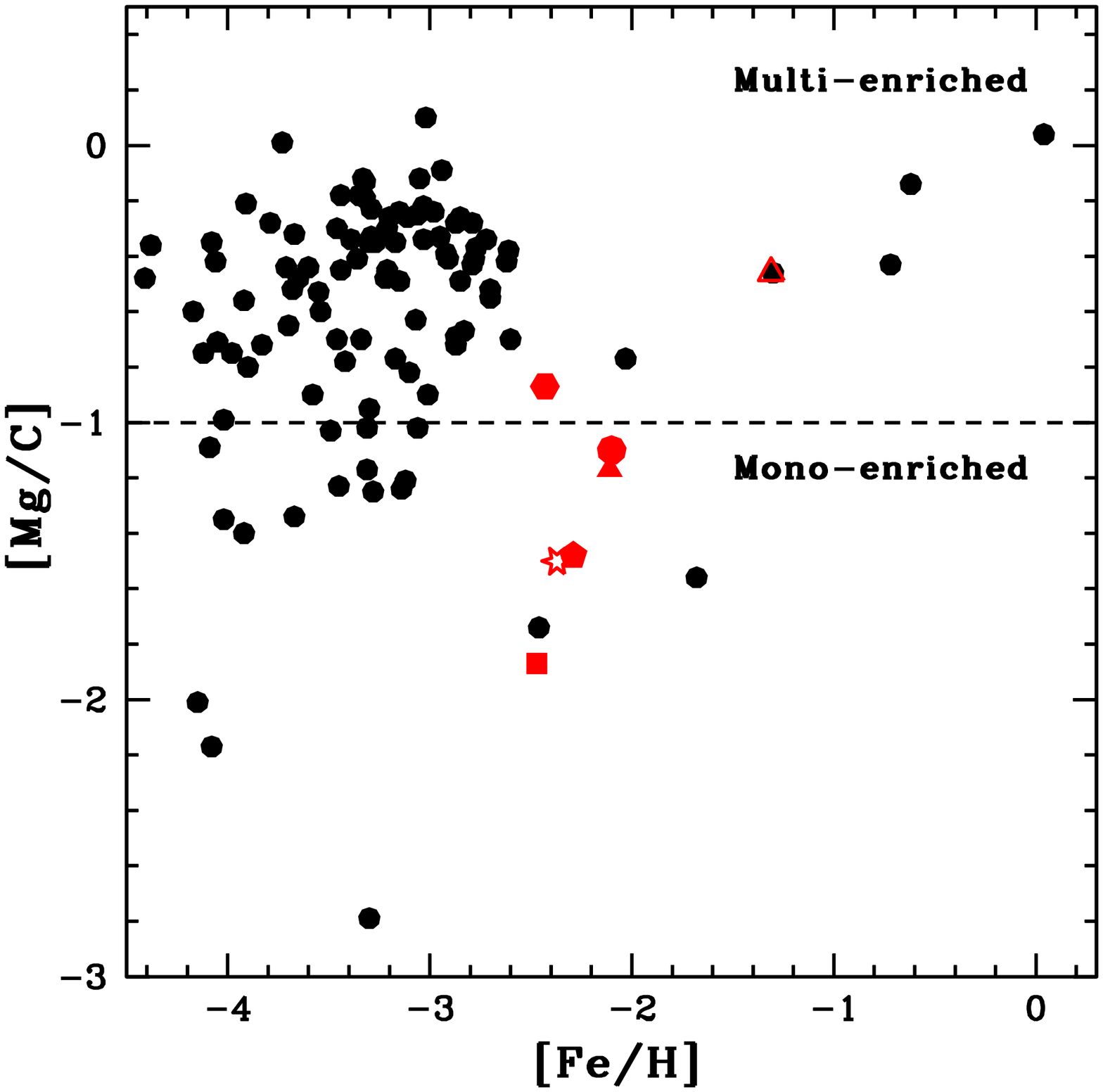}
\caption{Locations of programme stars in [Mg/C] vs. [Fe/H] plot. Symbols used for the programme stars are same as in Figure \ref{fig7}. Filled circles represent the stars from literature \citep{roederer.2014AJ....147..136R,plez.cohen.2005A&A...434.1117P,christlieb.2004A&A...428.1027C,hansen.2014ApJ...787..162H,frebel.2018ARNPS..68..237F,yong.2013ApJ...762...26Y,bonifacio.2015A&A...579A..28B,bessel.2015ApJ...806L..16B,lucatello.2003AJ....125..875L,barklem.2005A&A...439..129B,cohen.2006AJ....132..137C,aoki.2007ApJ...655..492A,goswami.2006MNRAS.372..343G,karinkuzhi.2015MNRAS.446.2348K,purandardas.2019aMNRAS.486.3266P,purandardas.2019bBSRSL..88..207P,partha.2021A&A...649A..49G,shejeela.2021MNRAS.502.1008S}}
\label{fig14}
\end{figure}

 Elemental abundance ratios such as [Mg/C], [Sc/Mn], [C/Cr] and [Ca/Fe] can also be used to examine  whether the star is polluted by a single event or several pollution events. \cite{hartwig.2018MNRAS.478.1795H} presented a novel diagnostic to identify stars that are formed from the gas enriched by only one previous supernova based on their cosmological models. They showed that [Mg/C] $<$ $-1.0$, [Sc/Mn] $<$ 0.50, [C/Cr] $>$ 0.50 and [Ca/Fe] $>$ 2 represent  mono-enrichment of a star. As an example we have shown the locations of our programme stars in the [Mg/C] vs. [Fe/H] diagram (Figure \ref{fig14}). All the programme stars except HE 1523$-$1155 fall in the region representing mono-enrichment. \cite{hansen.2019A&A...623A.128H} show that some of the CEMP-s stars in their sample are also mono-enriched. The estimated value of [Ca/Fe] in all our programme stars show that they are all multi-enriched as expected for CEMP-s and CEMP-r/s stars. The value of [Sc/Mn] measured in all the programme stars except HE 1425$-$2052 show that they are all mono-enriched. The estimated value of [C/Cr] ratio in the programme stars also show that they are mono-enriched except HE 1428$-$1950.  Various diagnostic elemental abundance ratios estimated in each of the programme star show that their surface composition is both mono- and multi- enriched which does not fit with the results from Hartwig et al. (2018). This indicates that the different diagnostic elements in our programme stars are not originated from the previous supernova and it may be attributed to their binary companion.

\subsection{Mixing diagnostic}
\begin{figure}
\centering
\includegraphics[width=12cm,height=11cm]{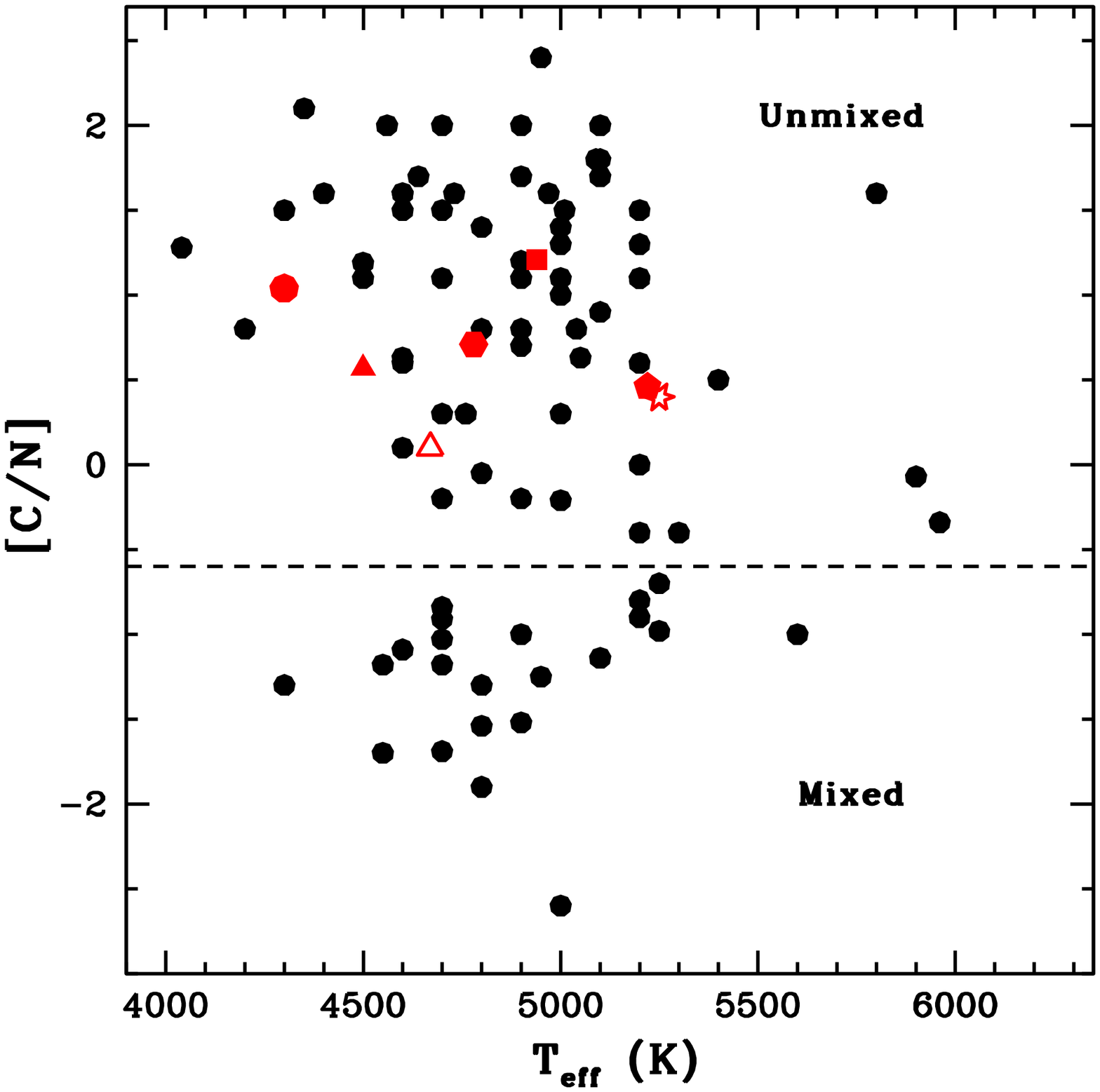}
\caption{Positions of the programme stars in the [C/N] vs. T$_{eff}$ diagram. Symbols used for the programme stars are same as in Figure \ref{fig7}. Filled circles represent the stars from literature \citep{spite.2006A&A...455..291S,aoki.2007ApJ...655..492A,goswami.2016MNRAS.455..402G,hansen.2016aA&A...586A.160H,hansen.2019A&A...623A.128H,purandardas.2019aMNRAS.486.3266P,purandardas.2019bBSRSL..88..207P,partha.2021A&A...649A..49G,shejeela.2021MNRAS.502.1008S}}
\label{fig15}
\end{figure}

\begin{figure}
\centering
\includegraphics[width=12cm,height=11cm]{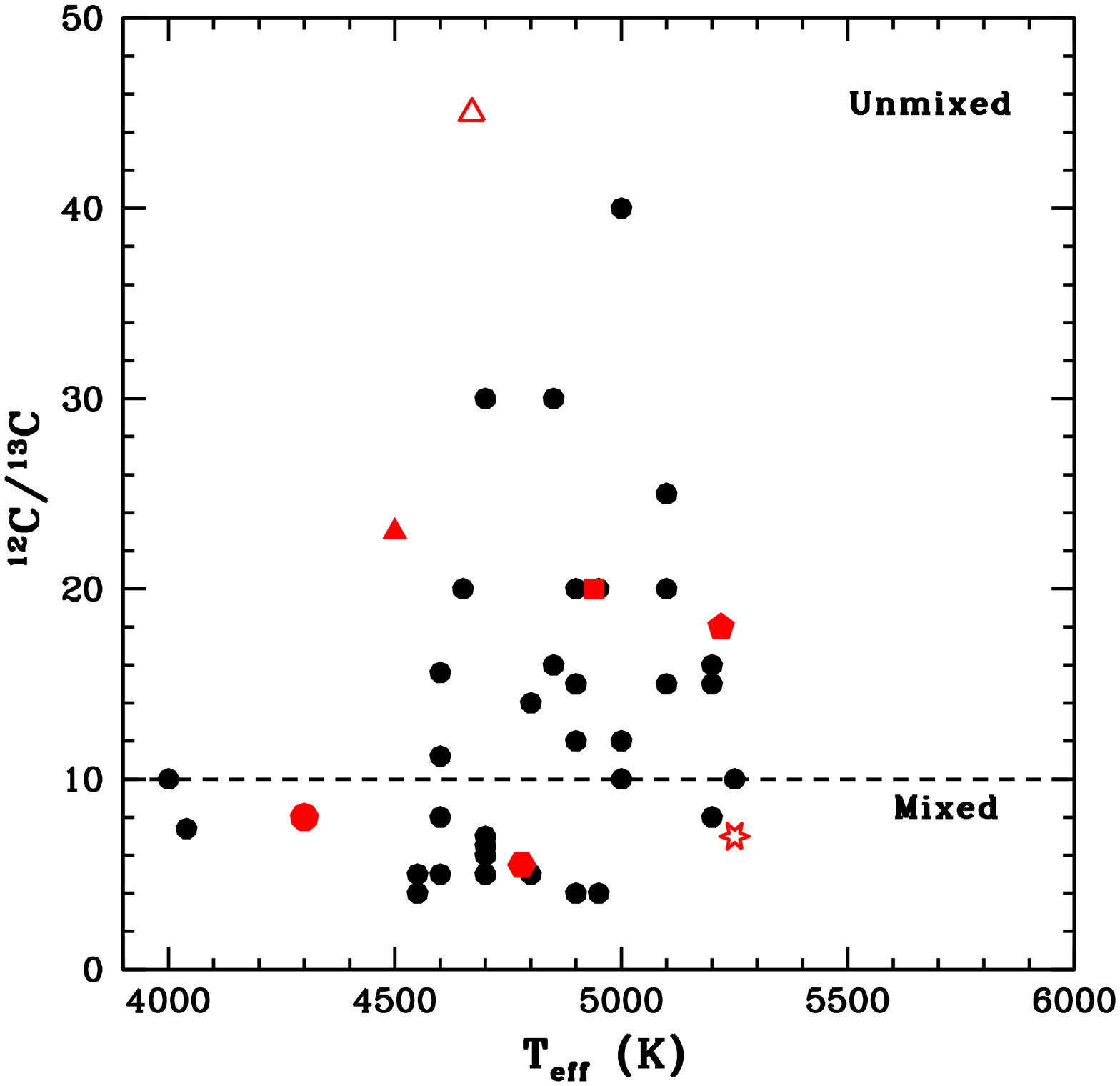}
\caption{Positions of the programme stars in $^{12}$C/$^{13}$C vs. T$_{eff}$ diagram. Symbols used for the programme stars are same as in Figure \ref{fig7}. Open circles represent the stars from \cite{spite.2006A&A...455..291S} and \cite{aoki.2007ApJ...655..492A}}
\label{fig16}
\end{figure}

As all the program stars are luminous objects with log(L/L$_{\odot}$) $>$ 2.50, an extra mixing can happen and modify the chemical compositions of the stars \citep{gratton.2000A&A...354..169G,spite.2005A&A...430..655S}. We have therefore examined any possibility of internal mixing processes that could have modified the observed surface compositions of our programme stars using estimates of [C/N] ratios and carbon isotopic ratios. None of the programme stars show any signatures of internal mixing based on [C/N] ratios (Figure \ref{fig15}). \cite{spite.2006A&A...455..291S} noted that [C/N] ratio is not a clean indicator of mixing as the abundances of carbon and nitrogen in the interstellar medium show large variations. In such cases, carbon isotopic ratio can be used as a good indicator of mixing as it is high in primordial matter and it is insensitive to the choice of atmospheric parameters for the stars \citep{spite.2006A&A...455..291S}. We could determine $^{12}C$/$^{13}$C ratio  in all our programme stars and the estimated values of $^{12}$C/$^{13}$C show that HE 1425$-$2052, HE 1523$-$1155 and HE 1528$-$0409 are well mixed objects (Figure \ref{fig16}).

\section{Conclusion}\label{sec:conclusions}

Results from follow-up high resolution spectral  analysis  of  seven potential CH star candidates from \cite{goswami.2005MNRAS.359..531G} and \cite{goswami.2010MNRAS.402.1111G} are presented.  Our analysis shows one object to be  metal-poor and the remaining six are very metal-poor stars. For the objects HE 0110$-$0406, HE 1425$-$2052 and HE 1428$-$1950, this work presents for the first time a detailed abundance analysis study. Our analysis shows that the objects HE 1447+0102 and HE 1523$-$1155 to be CEMP-r/s stars. The abundances of neutron-capture elements in HE 1447+0102 and HE 1523$-$1155 show good match with the yields from the i-process nucleosynthesis at a neutron density n $\sim$ 10$^{13}$. Based on our analysis, the objects HE 1425$-$2052, HE 1429$-$0551 and HE 1528$-$0409 are found to be CEMP-s stars. The object HE 1428$-$1950 could not be classified based on CEMP stars classification scheme as we could not estimate the abundance of Ba in this object.  

\par The locations of the programme stars in the absolute carbon abundance, A(C) vs. [Fe/H] diagram show that, they all belong to Group I objects. As most of the Group I objects are  known as binaries, the locations of the  stars among the Group I objects may indicate binarity. We note that, the object HE 1523$-$1155 in our sample  is a confirmed binary \citep{hansenetal.2016A&A...588A...3H}. 

\par The estimated [O/Fe] values indicate  AGB stars as  possible progenitors of the programme stars except for HE 1429$-$0551 and HE 1447+0102. For these two objects, the estimated oxygen abundance characterises  fast rotating massive star progenitors. 

\par Other useful indicators of the formation sources of the stars are [Sr/Ba] and [hs/ls] ratios. The estimated [Sr/Ba] ratio in HE 0110$-$0406, HE 1425$-$2052, HE 1447+0102 and HE 1523$-$1155 points at  AGB  stars as possible progenitors. The estimated [hs/ls] ratio in HE 0110$-$0406, HE 1425$-$2052, HE 1428$-$1950 and HE 1523$-$1155 also  indicate  AGB stars as  possible progenitors. 

\par We have also studied  the chemical enrichment histories of the programme stars based on various other elemental abundance ratios such as [Mg/C], [Sc/Mn] and [C/Cr]. While HE 1425$-$2052 is found to be multi-enriched, remaining stars are all found to be mono-enriched based on [Sc/Mn] ratio. The estimated [C/Cr] ratio in the programme stars also  shows that they are  mono-enriched except for HE 1428$-$1950. All the stars  are also found to be mono-enriched based on [Mg/C] ratio, except for HE 1523$-$1155. The estimated Ca  abundance of the stars show that their surface composition is enriched by multiple pollution events.

\par The programme stars are  luminous with log(L/L$_{\odot}$) $>$ 2. 
In order to understand whether any internal mixing processes have modified their surface compositions, we have looked  for  signatures of  internal mixing in our programme stars based on the estimates of the [C/N] and $^{12}$C/$^{13}$C ratios. While the [C/N] ratio estimated in the programme stars show that they are all unmixed, the estimated carbon isotopic ratios show that HE 1425$-$2052, HE 1523$-$1155 and HE 1528$-$0409 are well mixed. 

\par Kinematic analysis shows that the objects HE 0110$-$0406 and HE 1447+0102 to be thick disk objects and the remaining objects belong to the halo population of the Galaxy.
   
\par The estimates of abundance ratios  presented in this paper  can be used for theoretical studies to constrain the physics and the nucleosynthesis occurring in low-mass AGB stars. Abundance results of CEMP-r/s stars will also provide observational constrains for  i-process nucleosynthesis which is  believed to be the possible source of the origin of the peculiar chemical abundance patterns  observed in CEMP-r/s stars.
\section{ACKNOWLEDGEMENT}
Funding from the DST SERB project EMR/2016/005283 is gratefully acknowledged. We are thankful to Melanie Hampel for providing us with the i-process yields in the form of number fractions, and Partha Pratim Goswami for helping us to generate the model fits. We thank the referee for many constructive suggestions and useful comments on the manuscript which improved this paper.This work made use of the SIMBAD astronomical database, operated at CDS, Strasbourg, France, the NASA ADS, USA and data from the European Space Agency (ESA) mission Gaia (\url{https://www.cosmos.esa.int/gaia}), processed by the Gaia Data Processing and Analysis Consortium (DPAC, \url{https://www.cosmos.esa.int/web/gaia/dpac/consortium}). 

\section{Data Availability}
The data underlying this article will be shared on reasonable request to the authors.

\bibliography{cemp-s}{}
\bibliographystyle{aasjournal}

\appendix
\restartappendixnumbering
\section{\bf Lines used for deriving elemental abundances}

{\footnotesize
\begin{table*}
\caption{Lines used for deriving elemental abundances in our programme stars}
\resizebox{\textwidth}{!}{\begin{tabular}{ccccccccccc}
\hline 
Wavelength(\AA) & Element & $E_{low}$(eV) & log gf &HE 0110-0406&HE 1425-2052&HE 1428-1950&HE 1429-0551&HE 1447+0102&	HE 1523-1155&	HE 1528-0409\\
\hline
5682.633	&	Na I	&	2.102	&	-0.700	&	79.3(5.53)	&	50.1(4.75)	&	-	&	-	&	-	&	-	&	-	\\
5889.951	&	    	&	0.000	&	0.100	&	-	        &	-	        &	-	&	192.2(4.63)	&	162.5(4.56)	&	-	&	130.6(4.06)	\\
5895.920	&		&	0.000	&	-0.200	&	-	&	-	&	-	&	-	&	148.0(4.64)	&	180.2(4.53)	&	121.5(4.17)	\\
5172.684	&	Mg I	&	2.711	&	-0.402	&	-	&	-	&	-	&	172.2(5.55)	&	167.6(5.75)	&	-	&	-	\\
5528.405	&		&	4.346	&	-0.620	&	-	&	-	&	-	&	-	&	81.7(6.02)	&	101.5(6.13)	&	70.0(5.80)	\\
5711.088	&		&	4.346	&	-1.833	&	73.3(6.56)	&	47.0(5.83)	&	-	&	-	&	-	&	-	&	-	\\
4318.652	&	Ca I	&	1.899	&	-0.208	&	-	&	-	&	-	&	41.0(3.82)	&	-	&	59.4(4.11)	&	-	\\
5512.980	&		&	2.932	&	-0.290	&	47.4(4.85)	&	-	&	-	&	-	&	-	&	-	&	-	\\
5581.965	&		&	2.523	&	-0.710	&	-	&	58.5(4.50)	&	-	&	-	&	-	&	-	&	-	\\
5588.749	&		&	2.525	&	0.210	&	-	&	-	&	122.4(4.84)	&	35.8(3.91)	&	42.0(4.21)	&	59.6(4.28)	&	50.3(4.41)	\\
5590.114	&		&	2.521	&	-0.710	&	68.7(5.12)	&	-	&	-	&	-	&	-	&	-	&	-	\\
5857.451	&		&	2.932	&	0.230	&	-	&	107.1(4.68)	&	-	&	-	&	22.7(4.18)	&	35.5(4.23)	&	20.5(4.13)	\\
6102.723	&		&	1.879	&	-0.890	&	100.1(4.98)	&	119.9(4.52)	&	88.2(4.55)	&	25.0(4.01)	&	35.7(4.47)	&	46.0(4.32)	&	30.6(4.38)	\\
6162.173	&		&	1.899	&	0.100	&	-	&	-	&	-	&	-	&	66.4(4.15)	&	95.2(4.38)	&	70.7(4.29)	\\
6169.563	&		&	2.525	&	-0.270	&	-	&	77.6(4.25)	&	-	&	-	&	-	&	-	&	-	\\
6439.075	&		&	2.525	&	0.470	&	142.9(5.14)	&	172.1(4.65)	&	-	&	-	&	64.4(4.40)	&	-	&	53.1(4.18)	\\
6471.662	&		&	2.525	&	-0.590	&	-	&	111.5(4.95)	&	-	&	-	&	-	&	-	&	-	\\
4415.557	&	Sc II	&	0.595	&	-0.640	&	-	&	142.1(1.38)	&	-	&	79.5(1.03)	&	78.3(1.38)	&	88.6(1.24)	&	-	\\
4431.352	&		&	0.605	&	-1.880	&	-	&	74.8(1.58)	&	52.6(1.09)	&	-	&	-	&	20.6(0.87)	&	-	\\
5239.813	&		&	1.455	&	-0.770	&	-	&	-	&	-	&	30.6(1.01)	&	45.6(1.59)	&	42.9(1.23)	&	36.7(1.45)	\\
5526.790	&		&	1.768	&	0.130	&	123.7(2.00)	&	-	&	-	&	-	&	-	&	-	&	-	\\
6245.637	&		&	1.507	&	-0.980	&	85.3(1.99)	&	65.8(1.55)	&	-	&	-	&	-	&	26.4(1.09)	&	-	\\
4555.484	&	Ti I	&	0.848	&	-0.488	&	83.9(3.75)	&	-	&	-	&	-	&	-	&	-	&	-	\\
4759.270	&		&	2.255	&	0.514	&	-	&	-	&	-	&	-	&	-	&	-	&	-	\\
4820.410	&		&	1.500	&	-0.441	&	-	&	-	&	-	&	-	&	-	&	-	&	-	\\
4840.874	&		&	0.899	&	-0.509	&	72.8(3.58)	&	113.0(3.30)	&	-	&	-	&	-	&	-	&	-	\\
5024.844	&		&	0.818	&	-0.602	&	88.2(3.80)	&	-	&	-	&	-	&	-	&	-	&	-	\\
5210.385	&		&	0.047	&	-0.884	&	-	&	-	&	-	&	35.4(2.75)	&	-	&	52.1(2.85)	&	-	\\
5282.376	&		&	1.052	&	-1.300	&	37.0(3.92)	&	-	&	-	&	-	&	-	&	-	&	-	\\
5922.109	&		&	1.046	&	-1.466	&	-	&	23.1(3.11)	&	-	&	-	&	-	&	-	&	-	\\
4443.794	&	Ti II	&	1.080	&	-0.700	&	-	&	-	&	-	&	-	&	86.6(2.63)	&	-	&	-	\\
4468.507	&		&	1.130	&	-0.600	&	-	&	-	&	-	&	113.3(2.94)	&	-	&	115.2(2.94)	&	-	\\
4470.857	&		&	1.164	&	-2.280	&	-	&	-	&	78.2(2.95)	&	-	&	-	&	-	&	-	\\
4563.761	&		&	1.221	&	-0.960	&	-	&	-	&	-	&	-	&	62.1(2.33)	&	-	&	71.4(2.66)	\\
4583.409	&		&	1.164	&	-2.720	&	-	&	62.9(3.36)	&	-	&	-	&	-	&	-	&	-	\\
4764.526	&		&	1.236	&	-2.770	&	-	&	-	&	-	&	-	&	-	&	22.0(2.95)	&	-	\\
5185.913	&		&	1.892	&	-1.350	&	-	&	68.4(2.94)	&	-	&	50.5(2.88)	&	20.2(2.40)	&	46.8(2.80)	&	-	\\
5188.680	&		&	1.580	&	-1.210	&	-	&	-	&	-	&	-	&	-	&	-	&	59.7(2.91)	\\
5226.543	&		&	1.565	&	-1.300	&	-	&	-	&	154.5(3.39)	&	68.0(2.80)	&	38.3(2.47)	&	78.8(3.03)	&	54.5(2.86)	\\
5418.751	&		&	1.581	&	-1.999	&	-	&	-	&	-	&	-	&	-	&	-	&	-	\\
4379.230	&	V I	&	0.300	&	0.580	&	-	&	-	&	-	&	-	&	48.2(2.19)	&	-	&	-	\\
4864.731	&		&	0.017	&	-0.960	&	-	&	-	&	127.7(3.36)	&	-	&	-	&	-	&	-	\\
5703.575	&		&	1.050	&	-0.211	&	-	&	-	&	87.5(3.29)	&	-	&	-	&	-	&	-	\\
5737.059	&		&	1.063	&	-0.740	&	47.8(3.50)	&	25.3(2.46)	&	-	&	-	&	-	&	-	&	-	\\
6119.523	&		&	1.063	&	-0.320	&	-	&	70.3(2.63)	&	-	&	-	&	-	&	-	&	-	\\
6216.354	&		&	0.275	&	-1.290	&	66.1(3.27)	&	39.2(2.11)	&	-	&	-	&	-	&	-	&	-	\\
6251.827	&		&	0.286	&	-1.340	&	-	&	-	&	88.7(3.34)	&	-	&	-	&	-	&	-	\\
4652.157	&	Cr I	&	1.003	&	-1.030	&	-	&	-	&	178.8(4.92)	&	-	&	-	&	-	&	-	\\
4829.371	&		&	2.544	&	-0.810	&	-	&	-	&	52.2(4.70)	&	-	&	-	&	-	&	-	\\
4870.797	&		&	3.079	&	0.050	&	44.2(4.40)	&	-	&	-	&	-	&	-	&	-	&	-	\\
5296.691	&		&	0.982	&	-1.400	&	-	&	-	&	-	&	-	&	-	&	59.7(3.84)	&	-	\\
5345.796	&		&	1.003	&	-0.980	&	134.7(4.41)	&	-	&	-	&	-	&	-	&	67.7(3.62)	&	-	\\
5348.315	&		&	1.003	&	-1.290	&	98.5(4.06)	&	108.5(3.28)	&	-	&	23.8(3.18)	&	-	&	-	&	-	\\
4848.235	&	Cr II	&	3.864	&	-1.140	&	-	&	-	&	112.8(4.79)	&	-	&	20.6(3.72)&	27.4(3.64)	&	-	\\
4451.586	&	Mn I	&	2.888	&	0.278	&	-	&	-	&	105.3(4.26)	&	-	&	-	&	-	&	-	\\
4470.144	&		&	2.941	&	-0.444	&	-	&	30.0(3.31)	&	-	&	-	&	-	&	-	&	-	\\
4739.087	&		&	2.941	&	-0.490	&	64.3(4.40)	&	-	&	-	&	-	&	-	&	-	&	-	\\
4765.846	&		&	2.941	&	-0.080	&	-	&	39.8(3.07)	&	60.2(3.82)	&	-	&	-	&	-	&	-	\\
4783.427	&		&	2.298	&	0.042	&	-	&	-	&	121.7(3.82)	&	-	&	-	&	-	&	-	\\
4823.524	&		&	2.319	&	0.144	&	-	&	-	&	131.9(3.90)	&	-	&	-	&	-	&	-	\\
4792.846	&	Co I	&	3.252	&	-0.067	&	25.9(3.68)	&	-	&	-	&	-	&	-	&	26.9(3.86)	&	-	\\
4813.467	&		&	3.216	&	0.050	&	38.6(3.77)	&	19.7(2.97)	&	45.5(3.81)	& -	&	-	&	-	&	-	\\
5483.344	&		&	1.710	&	-1.490	&	59.9(3.73)	&	-	&	-	&	-	&	-	&	-	&	-	\\
4980.166	&	Ni I	&	3.606	&	0.070	&	-	&	-	&	-	&	-	&	-	&	42.5(4.31)	&	-	\\
6314.653	&		&	1.935	&	-1.770	&	-	&	-	&	106.9(4.41)	&	-	&	-	&	-	&	-	\\
6643.629	&		&	1.676	&	-2.300	&	129.9(5.36)	&	125.5(4.41)	&	88.2(4.35)	&	21.3(4.01)	&	22.9(4.37)	&	33.2(4.08)	&	-	\\
6767.768	&		&	1.826	&	-2.170	&	-	&	138.7(4.63)	&	-	&	-	&	-	&	36.9(4.20)	&	-	\\
4810.528	&	Zn I&	4.077	&	-0.137	&	56.0(2.84)	&	61.4(2.88)	&	78.2(2.95)	&	-	&	-	&	35.6(2.54)	&	-	\\
4607.327	&	Sr I	&	0.000	&	-0.570	&	63.0(2.49)	&	114.1(2.46)	&	37.9(1.97)	&	-	&	27.9(2.51)	&	-	&	-	\\
\hline
\end{tabular}}
\label{tableA1}
\end{table*}
}

{\footnotesize
\begin{table*}
\resizebox{\textwidth}{!}{\begin{tabular}{ccccccccccc}
\hline 
Wavelength(\AA) & Element & $E_{low}$(eV) & log gf &HE 0110-0406&HE 1425-2052&HE 1428-1950&HE 1429-0551&HE 1447+0102&	HE 1523-1155&	HE 1528-0409\\
\hline
4883.684	&	Y II	&	1.084	&	0.070	&	-	&	171.7(1.51)	&	142.3(1.30)	&	-	&	-	&	67.8(0.44)	&	-	\\
5087.416	&		&	1.084	&	-0.170	&	-	&	-	&	-	&	-	&	-	&	-	&	-	\\
6613.733	&		&	1.748	&	-1.110	&	80.7(2.20)	&	40.9(1.46)	&	-	&	-	&	-	&	-	&	-	\\
4130.645	&	Ba II	&	2.722	&	0.680	&	-	&	-	&	-	&	-	&	-	&	71.7(1.63)	&	-	\\
5853.668	&		&	0.604	&	-1.000	&	145.1(1.41)	&	-	&	-	&	125.0(1.74)	&	131.7(2.36)	&	-	&	106.3(1.99)	\\
6141.713	&		&	0.703	&	-0.076	&	-	&	-	&	-	&	163.9(1.49)	&	-	&	168.2(1.35)	&	180.6(2.16)	\\
6496.897	&		&	0.604	&	-0.377	&	-	&	-	&	-	&	-	&	-	&	-	&	154.7(2.04)	\\
4322.503	&	La II	&	0.173	&	-1.120	&	-	&	-	&	-	&	40.0(0.33)	&	47.9(0.90)	&	36.4(0.17)	&	39.8(0.75)	\\
4662.498	&		&	0.000	&	-1.240	&	-	&	149.7(1.34)	&	-	&	-	&	-	&	-	&	-	\\
4748.726	&		&	0.926	&	-0.860	&	-	&	101.8(1.43)	&	-	&	-	&	39.7(1.20)	&	-	&	-	\\
4921.776	&		&	0.244	&	-0.680	&	-	&	-	&	-	&	69.4(0.60)	&	68.9(1.07)	&	64.9(0.40)	&	53.0(0.67)	\\
5259.379	&		&	0.173	&	-1.760	&	-	&	143.4(1.79)	&	-	&	-	&	-	&	-	&	-	\\
5303.528	&		&	0.321	&	-1.430	&	-	&	124.9(1.39)	&	-	&	-	&	-	&	20.0(0.13)	&	-	\\
6320.376	&		&	0.172	&	-1.610	&	-	&	-	&	-	&	-	&	38.0(0.94)	&	-	&	-	\\
6390.477	&		&	0.321	&	-1.450	&	-	&	-	&	-	&	-	&	-	&	-	&	20.4(0.56)	\\
4336.244	&	Ce II	&	0.703	&	-0.564	&	-	&	-	&	-	&	25.6(0.81)	&	-	&	-	&	-	\\
4418.780	&		&	0.863	&	0.177	&	-	&	-	&	68.2(0.44)	&	-	&	-	&	-	&	-	\\
4427.916	&		&	0.535	&	-0.460	&	-	&	-	&	59.7(0.52)	&	-	&	-	&	-	&	28.2(0.96)	\\
4523.075	&		&	0.516	&	-0.304	&	-	&	-	&	-	&	-	&	-	&	47.6(0.74)	&	-	\\
4539.745	&		&	0.327	&	-0.459	&	-	&	-	&	-	&	-	&	-	&	54.5(0.84)	&	50.6(1.31)	\\
4544.953	&		&	0.417	&	-0.974	&	-	&	-	&	-	&	-	&	31.3(1.34)	&	-	&	-	\\
4562.359	&		&	0.477	&	0.081	&	-	&	-	&	-	&	82.9(1.37)	&	-	&	55.8(0.51)	&	-	\\
4725.069	&		&	0.521	&	-1.204	&	62.5(1.70)	&	86.1(1.70)	&	-	&	20.1(1.04)	&	20.7(1.33)	&	-	&	-	\\
5330.556	&		&	0.869	&	-0.760	&	43.5(1.27)	&	-	&	40.2(0.76)	&	20.4(0.95)	&	30.7(1.54)	&	20.2(0.72)	&	-	\\
5556.941	&		&	1.813	&	-0.405	&	18.1(1.50)	&	-	&	-	&	-	&	-	&	-	&	-	\\
5219.045	&	Pr II	&	0.795	&	-0.240	&	-	&	-	&	-	&	23.6(-0.63)	&	-	&	-	&	-	\\
5259.728	&		&	0.633	&	0.080	&	-	&	-&	-	&	-	&	-	&	-	&	-	\\
5292.619	&		&	0.648	&	-0.300	&	71.8(0.88)	&	-	&	-	&	-	&	22.4(0.50)	&	-	&	20.5(0.38)	\\
5322.772	&		&	0.482	&	-0.315	&	75.4(0.76)	&	-	&	-	&	-	&	25.4(0.41)	&	-	&	26.0(0.48)	\\
5892.251	&		&	1.439	&	-0.352	&	-	&	-	&	-	&	-	&	-	&	-	&	-	\\
6278.676	&		&	1.196	&	-0.630	&	-	&	-	&	-	&	-	&	-	&	-	&	-	\\
4412.256	&	Nd II	&	0.063	&	-1.420	&	-	&	-	&	-	&	-	&	31.9(1.22)	&	-	&	-	\\
4446.384	&		&	0.204	&	-0.590	&	-	&	-	&	-	&	48.3(0.54)	&	52.7(1.09)	&	50.8(0.52)	&	-	\\
4451.563	&		&	0.380	&	-0.040	&	-	&	-	&	105.0(0.37)	&	75.4(0.97)	&	68.2(1.24)	&	-	&	-	\\
4516.346	&		&	0.320	&	-0.950	&	-	&	-	&	-	&	40.3(0.84)	&	46.5(1.39)	&	-	&	26.2(0.94)	\\
4797.153	&		&	0.559	&	-0.950	&	-	&	99.5(1.45)	&	-	&	-	&	-	&	25.0(0.66)	&	-	\\
4825.478	&		&	0.182	&	-0.860	&	-	&	-	&	69.3(0.25)	&	-	&	46.7(1.10)	&	-	&	-	\\
4859.026	&		&	0.320	&	-0.830	&	-	&	143.1(1.65)	&	61.7(0.29)	&	-	&	35.0(0.94)	&	50.6(0.82)	&	-	\\
5212.361	&		&	0.204	&	-0.870	&	93.0(1.24)	&	169.2(1.79)	&	-	&	46.6(0.67)	&	57.9(1.39)	&	42.2(0.48)	&	34.5(0.87)	\\
5255.506	&		&	0.204	&	-0.820	&	-	&	146.2(1.39)	&	-	&	53.3(0.76)	&	-	&	52.4(0.65)	&	26.7(0.62)	\\
5287.133	&		&	0.744	&	-1.300	&	-	&	91.0(1.84)	&	-	&	-	&	-	&	-	&	-	\\
5293.163	&		&	0.822	&	-0.060	&	-	&	-	&	-	&	47.7(0.60)	&	64.8(1.48)	&	47.1(0.52)	&	39.8(0.87)	\\
5319.815	&		&	0.550	&	-0.210	&	-	&	-	&	-	&	56.3(0.62)	&	69.5(1.45)	&	60.4(0.64)	&	57.7(1.19)	\\
5356.967	&		&	1.264	&	-0.250	&	76.9(1.62)	&	-	&	-	&	-	&	27.9(1.20)	&	-	&	-	\\
5442.264	&		&	0.680	&	-0.910	&	81.0(1.61)	&	-	&	-	&	-	&	-	&	-	&	-	\\
5702.238	&		&	0.744	&	-0.770	&	81.4(1.51)	&	138.7(1.88)	&	-	&	-	&	37.1(1.33)	&	19.0(0.44)	&	-	\\
4220.661	&	Sm II	&	0.544	&	-1.114	&	-	&	-	&	-	&	-	&	24.0(0.99)	&	-	&	-	\\
4424.337	&		&	0.485	&	-0.260	&	-	&	-	&	-	&	-	&	-	&	45.5(0.16)	&	55.7(0.99)	\\
4434.318	&		&	0.378	&	-0.576	&	-	&	158.1(1.70)	&	59.5(-0.04)	&	40.3(0.29)	&	-	&	51.1(0.48)	&	-	\\
4499.475	&		&	0.248	&	-1.413	&	-	&	92.7(1.24)	&	-	&	-	&	-	&	20.0(0.33)	&	-	\\
4519.630	&		&	0.543	&	-0.751	&	-	&	-	&	-	&	-	&	38.0(0.94)	&	32.2(0.39)	&	-	\\
4566.202	&		&	0.333	&	-1.245	&	-	&	100.3(1.28)	&	-	&	-	&	38.7(1.22)	&	33.4(0.65)	&	-	\\
4615.444	&		&	0.543	&	-1.262	&	-	&	98.8(1.55)	&	-	&	-	&	-	&	-	&	-	\\
4674.593	&		&	0.184	&	-1.055	&	65.7(0.72)	&	-	&	-	&	-	&	-	&	-	&	-	\\
4726.026	&		&	0.333	&	-1.849	&	25.0(0.92)	&	-	&	-	&	-	&	-	&	-	&	-	\\
4791.580	&		&	0.104	&	-1.846	&	47.8(1.07)	&	119.5(1.79)	&	-	&	-	&	-	&	-	&	-	\\
4854.368	&		&	0.378	&	-1.873	&	20.3(0.87)	&	79.7(1.61)	&	-	&	-	&	-	&	-	&	-	\\
4129.730	&	Eu II	&	0.000	&	0.204	&	-	&	-	&	-	&	-	&	74.9(0.11)	&	-	&	22.3(-1.45)	\\
6437.640	&		&	1.319	&	-0.276	&	-	&	70.1(0.49)	&	-	&	-	&	22.4(0.25)	&	-	&	-	\\
6645.064	&		&	1.379	&	0.204	&	-	&	89.3(0.31)	&	64.6(-0.35)	&	-	&	-	&	-	&	-	\\
\hline
\end{tabular}}
The numbers in the  parenthesis in columns 5-11 give the derived abundances from the respective line.
log gf values are taken from  kurucz atomic line list (\url{https://www.cfa.harvard.edu}).
\end{table*}
}

\end{document}